\documentclass[aps,pre,groupedaddress,notitlepage]{revtex4}
\usepackage{graphicx,amsmath,amssymb,mathtools,gensymb,xr} 
\usepackage[usenames,dvipsnames]{xcolor}

\begin{document}
\title{A new look at effective interactions between microgel particles}
\author{Maxime J. Bergman$^1$}
\author{Nicoletta Gnan$^2$}
\author{Marc Obiols-Rabasa$^{1a}$}
\author{Janne-Mieke Meijer$^{1b}$}
\author{Lorenzo Rovigatti$^{2}$}
\author{Emanuela Zaccarelli$^{2}$}
\email[]{emanuela.zaccarelli@cnr.it}
\author{Peter Schurtenberger$^{1}$} 
\email[]{peter.schurtenberger@fkem1.lu.se}
\affiliation{$^1$ Division of Physical Chemistry, Department of Chemistry, Lund University, PO Box 124, SE-22100 Lund, Sweden.}
\affiliation{$^2$ CNR-ISC and Department of Physics, Sapienza University of Rome, Piazzale A. Moro $2$, $00185$ Roma, Italy}
\affiliation{$^a$ current address: CR Competence AB, Naturvetarev\"{a}gen 14, 22362 Lund, Sweden}
\affiliation{$^b$ current address: Department of Physics, University of Konstanz, PO Box 688, D-78457 Konstanz, Germany.}
\maketitle
%\date{} % Leave empty to omit a date
%\renewcommand{\maketitlehookd}{%
%\begin{abstract}
%\normalsize
%\vspace{-30pt} %Moves the abstract upwards to minimize gap after author affiliations

    {\bf Thermoresponsive microgels find widespread use as colloidal model systems, because their temperature-dependent size allows facile tuning of their volume fraction {\it in situ}. However, an interaction potential unifying their behavior across the entire phase diagram is sorely lacking. Here we investigate microgel suspensions in the fluid regime at different volume fractions and temperatures, and in the presence of another population of small microgels, combining confocal microscopy experiments and numerical simulations. We find that effective interactions between microgels are clearly temperature dependent. In addition, microgel mixtures possess an enhanced stability compared to hard colloid mixtures - a property not predicted by a simple Hertzian model. Based on numerical calculations we propose a multi-Hertzian model, which reproduces the experimental behaviour for all studied conditions. Our findings highlight that effective interactions between microgels are much more complex than usually assumed, displaying a crucial dependence on temperature and the internal core-corona architecture of the particles.}\\
   
%----------------------------------------------------------------------------------------
%\section*{Introduction}

Microgels are hybrid particles with a dual colloid-polymer nature, belonging to the class of so-called soft colloids\cite{Vlassopoulos:14, Vanderscheer:17}. A microgel consists of a mesoscopic cross-linked polymer network, which can deform, shrink or interpenetrate with another microgel\cite{Gaurasundar2017, Mohanty2017}. Often, as a result of the synthesis conditions, a particle possesses a denser core and a more loosely crosslinked corona\cite{Stieger:04,Rey:16}, which also includes so-called dangling ends\cite{Boon:17,Gnan:17}.
Microgels are considered smart colloidal materials: in response to external parameters such as temperature, pH, ionic strength, light or electric field (depending on the nature of the polymers) a particle is able to change its size as well as other connected properties such as the polarizability\cite{mohanty:16} or elasticity of the particle\cite{hashmi2009mechanical,bachman2015ultrasoft}. Thus, they are promising for and already employed in several applications, such as photonic crystals\cite{Reese:04, Serpe2012}, drug delivery systems\cite{Hamidi:08, Peppas:00, oh:08} or nanotechnologies\cite{fernandez:09}. In addition, thanks to their high tunability and to their softness, microgels represent ideal model systems to study phase transitions\cite{wang:12,hilhorst:11,peng:10,alsayed:05, Mohanty:15} and glass  or jamming transitions in dense colloidal dispersions\cite{Zhang:09, Caswell:13, Yunker:14}.

In the case of thermoresponsive microgels made of Poly(N-isopropylacrylamide) (PNIPAM), the soft colloids are swollen below the volume phase transition temperature (VPTT) of 32$\degree$C\cite{Heskins:68, pelton:00,pelton:86}. At temperatures $T >$ VPTT, the swollen microgel network collapses and expels a significant fraction of water\cite{pelton:00, Romeo:10}. Thus, temperature is readily used as a convenient parameter to control {\it in situ} the size and the volume fraction of microgel samples\cite{Heyes:09, Yunker:14,wang:12,hilhorst:11,peng:10,alsayed:05}. In doing so, however, one implicitly assumes that such a temperature change does not alter the effective interactions between microgels. Early studies have proposed to model effective interactions between swollen microgel particles (below the VPTT) in terms of a hard-sphere like potential, with a modified/effective hard sphere diameter\cite{Senff:99, Stieger:04, Wu:031}, whereas for $T >$ VPTT  microgels should behave as attractive spheres\cite{Wu:031}. Recent research shows that the interactions between swollen microgels can be more accurately reproduced by a soft Hertzian repulsion in the fluid region of the phase diagram\cite{Mohanty:14}, while brush-like models can be used for highly packed samples\cite{Scheffold:10, Romeo:13}.  All these different models point to the surprising fact that there is not yet a unifying picture which can describe microgels' interactions. Clearly there is the need to carefully characterize the interparticle potential under different experimental conditions across the entire phase space. Such a step is necessary not only for a correct use of microgel systems in their widespread applications but also from a fundamental point of view: only fully characterised systems should be used to work on open problems in condensed matter physics, such as glass transition and jamming.  

In this work, we investigate the effective interactions of microgels in a wide region of the fluid regime, and as a function of temperature for $T <$ VPTT, i.e. for swollen microgels. We study both one-component microgel suspensions and binary mixtures in which much smaller microgels are added, inducing an effective depletion on the large ones. For each state point, experimental structural and dynamical information was compared to its simulated counterpart. We confirm the applicability of a soft repulsive Hertzian interaction potential for the one-component system, even at elevated temperatures. However, the Hertzian model predicts instantaneous aggregation of the large microgels in the mixtures, which experience a depletion attraction. In contrast, all mixtures are stabilised by the core-repulsion of the microgels. Based on numerical calculations of the effective interaction potential, we develop a multi-Hertzian model, which ascribes a different elasticity to corona-corona interactions - reflecting the simple Hertzian interactions between microgels at moderate packing fractions - and core-corona interactions. The MH model captures the structure and the dynamical behavior of the studied binary mixtures at all 48 investigated different state points. Evidently, it is imperative to consider the variation of the interparticle potential upon changing temperature and, crucially, the internal structure of the microgels to correctly describe their behavior, particularly for conditions where microgels are forced together - for example, in electric field applications or in the dense glassy regime, which is most widely studied in the microgel literature. Furthermore, our results raise fundamental questions on the widespread practice to tune the volume fraction via a temperature change without accounting for the different nature of the system. \\
 
\textbf{Results\\}

%%%%COLLOID-ONLY%%%%%
%------------------------------------------
\textbf{Structure and dynamics of one-component microgel systems\\}  
  
\begin{figure*}[h!]
\centering
  \includegraphics{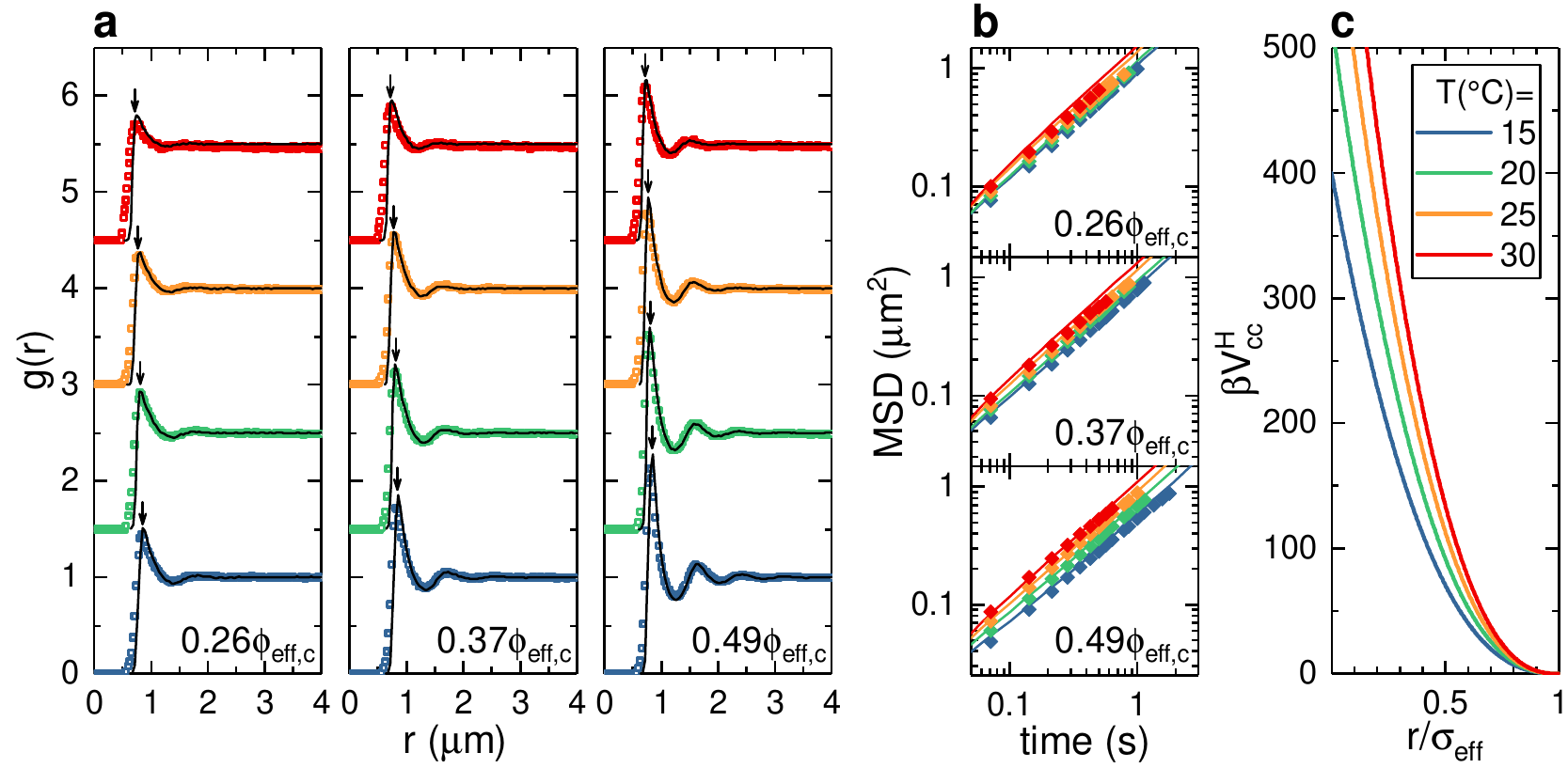}
  \caption{{\bf Temperature-dependent structure, dynamics and interaction potential of one-component microgel suspensions.}
Symbols indicate experimental data, solid lines represent simulations. Color legend applies to all panels. 
{\bf a)} Radial distribution functions $g(r)$ from simulations and experiments. Panels show data for $\phi^{15\degree \text{C}}_{\text{eff,c}} =$ 0.26, 0.37 and 0.49 (left to right) for $15\degree$C $ \leq T\leq 30\degree$C. Graphs are offset along $y$-axis for clarity. Downward pointing arrows indicate the hydrodynamic diameter of colloids at each $T$. 
{\bf b)} MSDs for $15\degree$C $ \leq T\leq 30\degree$C reconstructed from the $x,y$ trajectories, i.e. $\langle x^2+y^2\rangle$, with $\phi^{15\degree \text{C}}_{\text{eff,c}} =$ 0.26, 0.37 and 0.49 (from top to bottom).
 {\bf c)} Hertzian interaction potential at different temperatures: $U_{\text{cc}}=400, 520, 640, 760k_\text{B} T$ for $T=15,20,25,30\degree$C respectively. The distance is rescaled by $\sigma_{\text{eff}}=2R_\text{H}$.}
  \label{fig:coll_only}
  \end{figure*}

We start by analyzing the behavior of one-component microgel suspensions (also referred to as 'colloid' samples) with weight fractions wt\% $=$ 2.2, 3.3, 4.4. The radial distribution function ($g(r)$) and mean squared displacements (MSD) of the samples were measured at four different temperatures in the range 15-30$\degree$C. We find, as expected, that an increase in particle concentration leads to an increase in the structural correlations for all $T$ (Figure \ref{fig:coll_only}a). An increase in temperature is associated with the deswelling of particles, which is quantified by additional dynamic light scattering measurements (Supplementary Fig. \ref{fig:swelling}, Supplementary Note \ref{fig:swelling}), and thus to a decrease of the volume fraction of the sample. This leads to a reduced structural order as well as to the shift of the main peak of the $g(r)$ toward smaller distances. From the trajectories obtained with confocal scanning laser microscopy (CLSM) we also reconstruct the MSDs in a time window within the long-time diffusive regime of the microgels (Figure \ref{fig:coll_only}b). We find that increasing temperature speeds up particle diffusion, due to the reduction in volume fraction as well as to the faster thermal motion and to a reduction in solvent viscosity.

%%%%FIGURE 1%%%%%
 
In order to describe the experimental behavior we use the soft Hertzian-type repulsion which has been previously shown to accurately describe microgel interactions in the fluid phase at 15$\degree$C\cite{Paloli:12, Mohanty:14}. The (colloid-colloid) Hertzian potential $V^\text{H}_{\text{cc}}(r)= U_{\text{cc}} (1-r/\sigma_{\text{eff}})^{5/2}\theta(\sigma_{\text{eff}}-r)$ where $\theta(r)$ is the Heaviside step function, depends on two control parameters: the effective colloid diameter $\sigma_{\text{eff}}$ and the interaction strength at full overlap $U_{\text{cc}}$. We fix the former to be equal to $2 R_\text{H}$, where $R_\text{H}$ is the (experimentally determined) hydrodynamic radius of the particles at each considered $T$ (Supplementary Fig. \ref{fig:swelling}, Supplementary Note \ref{fig:swelling}). Next, we adjust the colloid volume fraction $\phi_{\text{eff,c}}$ at $T=$15$\degree$C around the value predicted by viscometry (see Methods, Supplementary Fig. \ref{fig:suppl-viscometry}, Supplementary Note \ref{fig:suppl-viscometry}) and we vary $U_{\text{cc}}$ until a good correspondence is found with the experimental $g(r)$s.

We find, in line with previous work with slightly different microgels\cite{Mohanty:14}, that the interaction strength at $T=$15$\degree$C is $U_{\text{cc}}=400k_\text{B} T$ and the three colloid packing fractions, that will serve as a basis also for the binary mixtures discussed later on, are  $\phi_{\text{eff,c}} =$ 0.26, 0.37 and 0.49 at 15$\degree$C. 
To model the variation in temperature, the volume fractions are changed according to the deswelling of the microgels (Supplementary Fig. \ref{fig:swelling}, Supplementary Note \ref{fig:swelling}), and again we vary the interaction strength $U_{\text{cc}}$ until the experimental radial distribution functions are well reproduced also at higher $T$. Consequently, we find that the numerical and experimental $g(r)$ are in good agreement for all investigated state points. In particular, the positions of all peaks is well-captured and the secondary peaks are quantitatively reproduced, while some deviations are observed close to the main peak. However, there is no systematic trend of such deviations with respect to packing fraction, as shown and described in Supplementary Fig.~\ref{fig:chi_2} and Supplementary Note \ref{fig:chi_2}. This suggests that the discrepancy is mostly driven by data noise. A systematic worsening of the agreement is found at $T=$30$\degree$C, where lower spatial resolution of the CLSM in the $z$-direction and the rapid Brownian motion of the particles leads to a reduction in the peak height and a broadening of the $g(r)$ data for values to the left of the first peak~\cite{Mohanty:14}.

The agreement between experimental and numerical data indicates that the Hertzian model is able to describe the structure of the swollen microgel system in the range of investigated packing fractions. Furthermore, we find that $U_{\text{cc}}$ approximately follows a linear dependence on temperature  (Supplementary Fig. \ref{fig:Ucd-linear}, Supplementary Note \ref{fig:Ucd-linear}). 
The fact that raising the temperature makes PNIPAM more hydrophobic might give rise to expectations that by increasing $T$ microgels become more and more attractive~\cite{Romeo:10,Wu:031}. By contrast, our results show that the Hertzian repulsion is an increasing rather than decreasing function of temperature, at least up to $30^\circ$C (Figure 1c), in agreement with static light scattering experiments previously obtained with microgels with a lower crosslink density~\cite{Wu:031}. Only close to the VPT and beyond, outside of the regime explored in this work, attractive interactions become dominant~\cite{Wu:031}. 

A further test of the Hertzian model can be made by comparing experimental and numerical MSDs. To this aim we use Brownian Dynamics (BD) simulations, which show that that the Hertzian model is also able to reproduce the variation of the MSDs (Figure \ref{fig:coll_only}b) with $T$ and $\phi_{\text{eff,c}}$ in the investigated regime. The direct comparison between the numerical and experimental self-diffusion coefficients is reported in Supplementary Fig.~\ref{fig:diff_coeff} and Supplementary Note \ref{fig:diff_coeff}.

%%%%%FIGURE 2%%%%%

\begin{figure*}[h!]
\centering
  \includegraphics{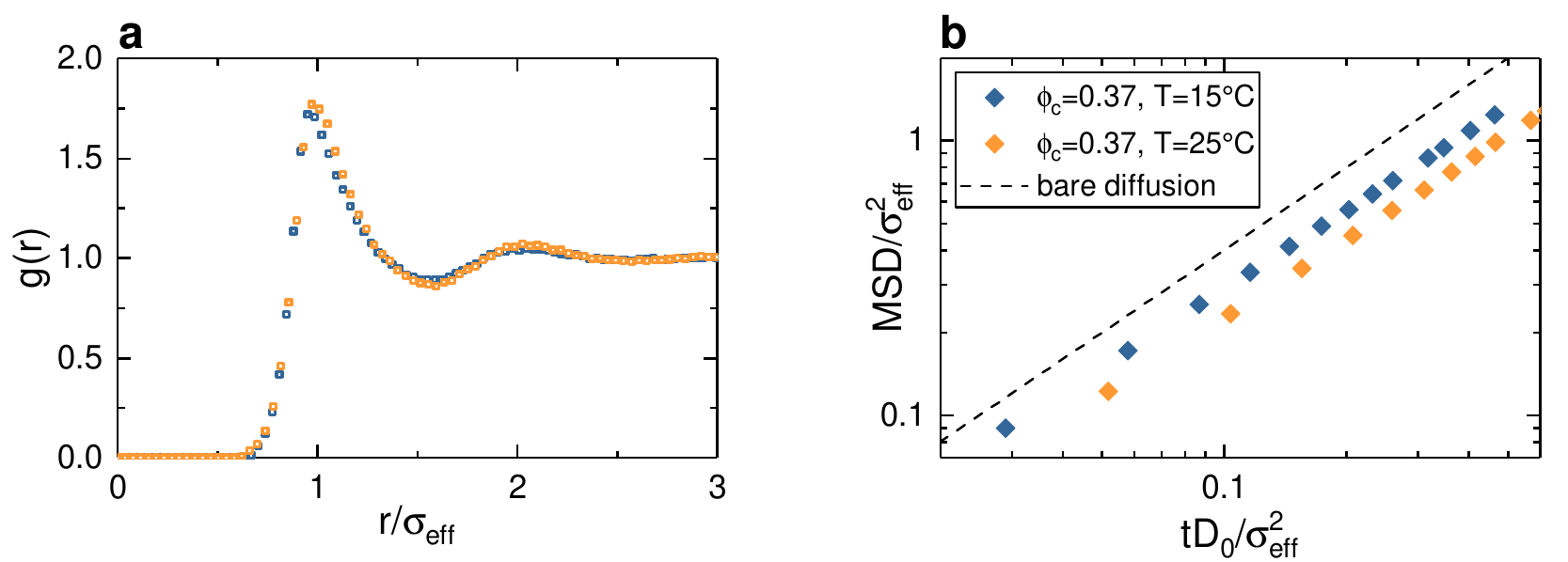}
 \caption{{\bf Structure and dynamics for two state points at different temperature with equivalent packing fraction.} Experimental data for samples with $\phi_{\text{eff,c}}= 0.37$ at two different temperatures ($T= 15$ and 25$\degree$C) corrected for size. Also shown in {\bf b)} is the predicted diffusion (dashed line). $D_0$ is the zero-colloid limit diffusion coefficient.}
  \label{fig:phi037}
\end{figure*}

It is particularly interesting to directly compare samples with the same effective volume fraction but at different temperatures, i.e. samples with unequal number density, as shown in Figure \ref{fig:phi037}. Comparing the experimental data for $g(r)$, we find a weak but detectable increase of the correlation with increasing temperature. Indeed, at the higher $T$ the first two peaks increase their height and shift towards larger separations, the in-between minimum deepens and a weak oscillation beyond $r/\sigma_{\rm eff} = 2$ appears. Not only the structure of the system is affected, but also the dynamics. After appropriately rescaling the data to take into account the different hydrodynamic radii and zero-colloid limit diffusion coefficients at each $T$, the experimental MSDs show a marked difference between the two samples: the $T=25\degree$C system is much slower than the one at lower temperature. Thus, at higher $T$ the system is more structured as well as slower, which is consistent with the stronger Hertzian repulsion that we have determined within our theoretical analysis. However, it is important to stress that the present evidence is based solely on experimental data and does not rely on the particular choice of any model. Our measurements clearly show that the interaction potential between the particles changes as a function of temperature, even within the swollen regime (15-30$\degree$C). 

To summarize, we have found that a temperature-dependent Hertzian model can be used to correctly capture both the structure and the dynamics of one-component microgel suspensions in the investigated temperature and packing fraction range. Importantly, the Hertzian repulsion is found to increase with temperature. These findings directly confirm the hypothesis that, by changing the temperature, not only the packing fraction is varied but also the interaction potential is considerably affected. This is particularly important for studies in which the temperature is used as a facile way to tune the effective volume fraction of soft microgels, where these temperature-dependent changes in interparticle interactions should be carefully considered.\\

\textbf{The Hertzian model poorly describes microgel mixtures\\} 

We now turn to analyze mixtures of large (colloid) and small (depletant) microgels. The very small size ratio $R_{\text{H,depletant}}/R_{\text{H,colloid}}$ changes very little, i.e. from 0.055 to 0.060, within the investigated temperature range (Supplementary Fig. \ref{fig:swelling}, Supplementary Note \ref{fig:swelling}).  In this framework, it is possible to derive an effective interaction potential for large microgels only, integrating out the small particles' degrees of freedom\cite{Likos:01}. The small microgels thus induce a depletion interaction between the colloids. 

\begin{figure*}[h!]
\centering
  \includegraphics{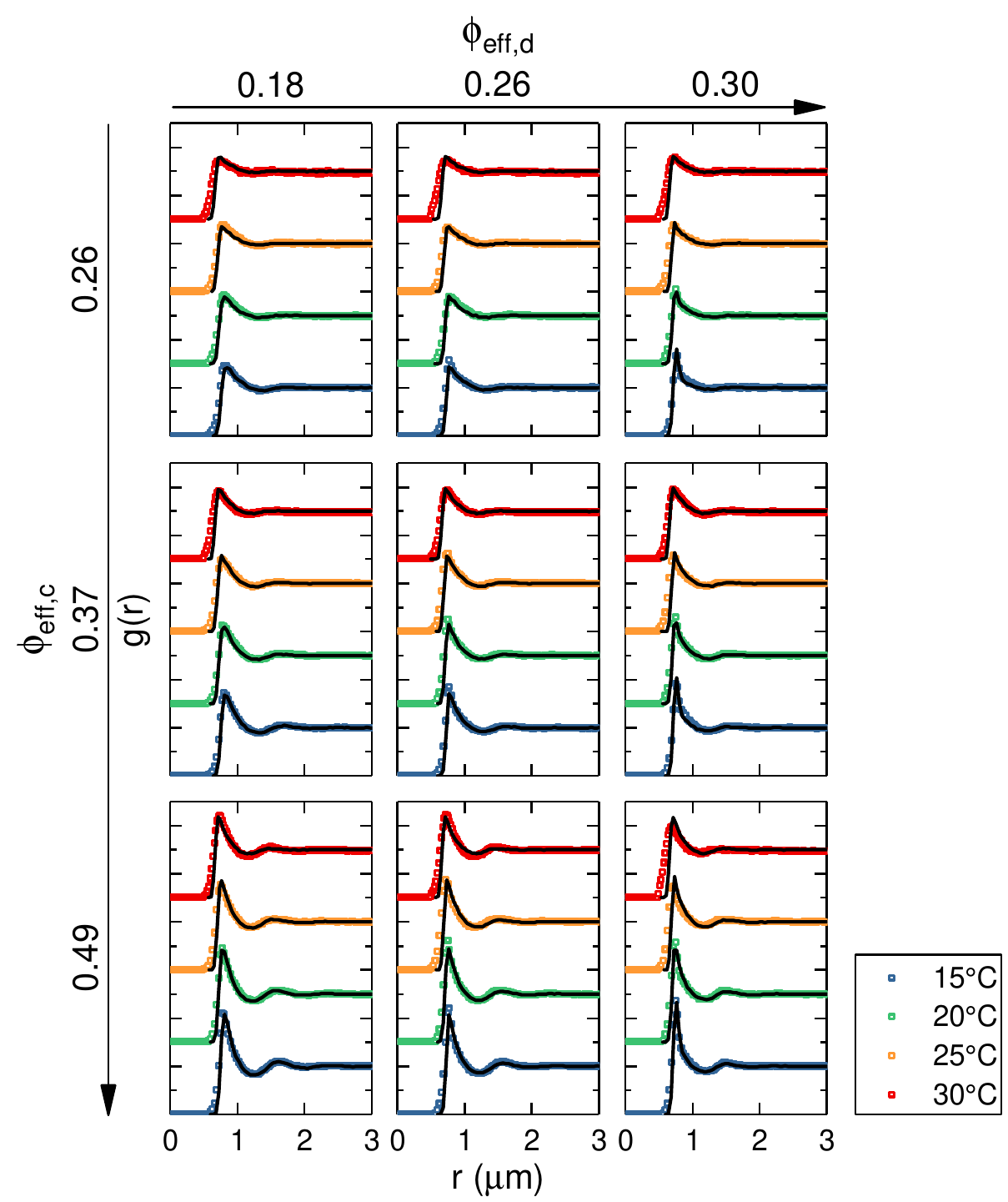}
  \caption{{\bf Experimental and numerical structural correlations for all investigated binary mixtures.} Experimental $g(r)$s (colored squares) are compared to numerical ones (solid lines) based on the multi-Hertzian model. Data for different samples are offset in $y$ for clarity. The color legend applies to the entire graph. Values of $\phi_{\text{eff,c}}$ and $\phi_{\text{eff,d}}$ at 15$\degree$C are given for each row and column, respectively.  For higher temperatures, the values of $\phi_{\text{eff,c}}$, $\phi_{\text{eff,d}}$ can be found in Table \ref{tab:phieff}. }
  \label{fig:T15-30}
  \end{figure*}

We investigate nine colloid-depletant mixtures with colloid wt\% $=$ 2.2, 3.3, 4.4 and depletant wt\% $=$ 0.26, 0.54, 0.81. We start by analysing  structural correlations (Figure \ref{fig:T15-30}). In the presence of depletants, the colloidal particles show an increased attraction: the first maximum of the $g(r)$ increases in height and becomes asymmetric. In addition, the nearest neighbor distance, characterised by the position of such maximum, decreases with the addition of depletants.  

%%%%%FIGURE 3%%%%%%

%%%%%FIGURE 4%%%%%%

\begin{figure}[h!]
\centering
  \includegraphics{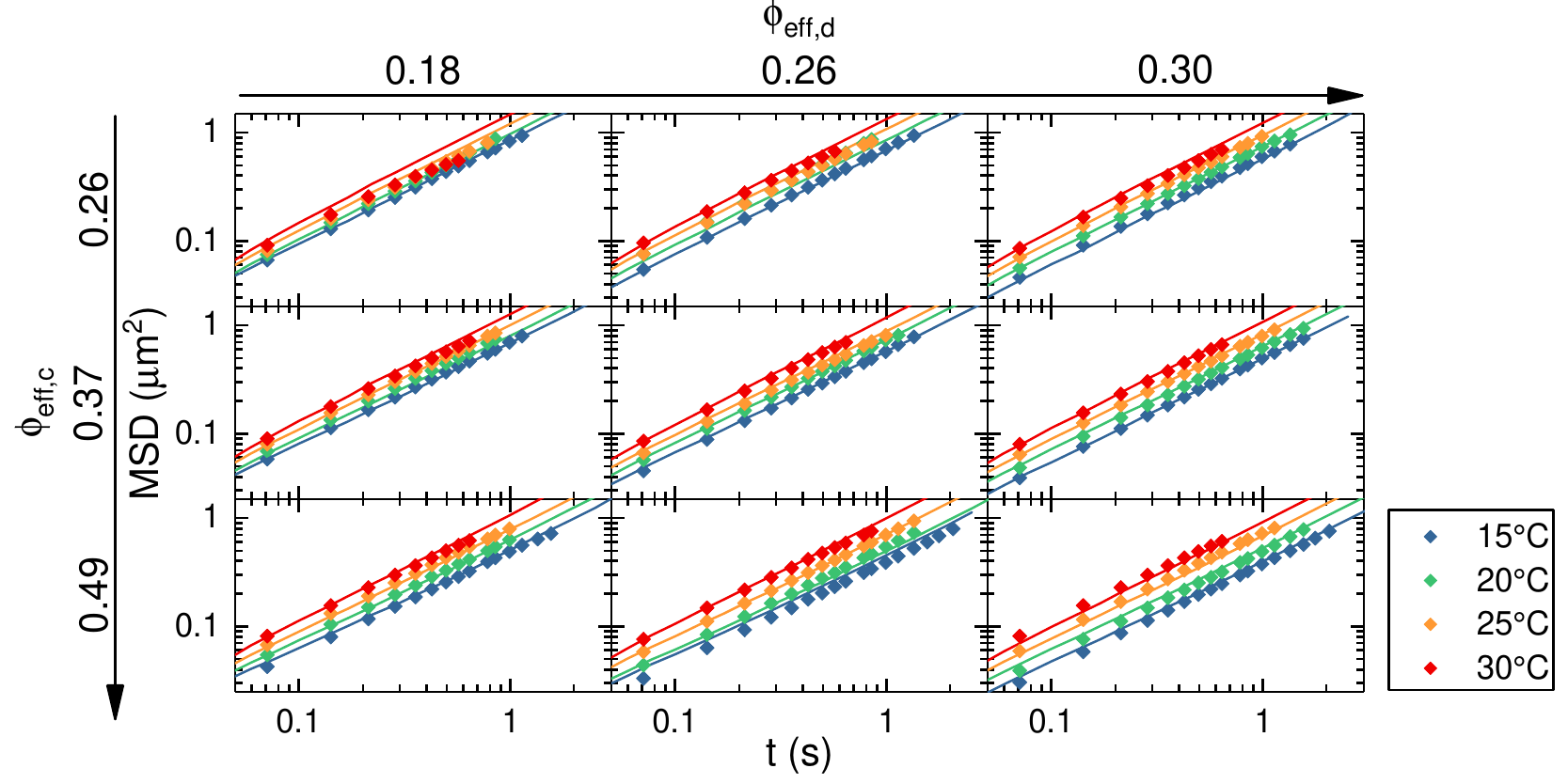}
  \caption{{\bf Experimental and numerical mean square displacements for all investigated state points}. Diamonds denote 2D experimental data ($\langle x^2+y^2\rangle$), while solid lines represent the corresponding simulation results based on the MH model.  The color legend applies to entire graph. Values of $\phi_{\text{eff,c}}$ and $\phi_{\text{eff,d}}$ at 15$\degree$C are given for each row and column, respectively.  For higher temperatures, the values of $\phi_{\text{eff,c}},\phi_{\text{eff,d}}$ can be found in Table \ref{tab:phieff}.  }
  \label{fig:msds}
\end{figure}

A striking result from the experiments is that the depletion attraction is not as strong as expected: all studied mixtures are surprisingly stable and fluid-like. Comparing with recent results for a binary mixture of hard spheres (colloids) and microgels (depletants), phase separation was observed well within the currently investigated range of depletant concentrations\cite{Bayliss:11}. Furthermore, theoretical models for depletion among soft particles would predict a strongly enhanced depletion attraction as compared to that occurring in corresponding hard-sphere systems\cite{Rovigatti:15}.  On the contrary, our experimental findings show that the effect of the depletion attraction is small. This is confirmed by the variation of the MSDs with added depletants (Figure \ref{fig:msds}): upon increasing depletant concentration we only observe a moderate slowing down of the diffusion. 

In order to describe the observed behavior, we start by modeling the binary mixtures at $T=15\degree$C using again the Hertzian model. Thus, the total interaction in the mixture amounts to $V_{\text{tot}}=V^\text{H}_{\text{cc}} + V^\text{H}_{\text{cd}} + V^\text{H}_{\text{dd}}$, where the three terms are the direct colloid-colloid interaction, the colloid-depletant interaction and the direct depletant-depletant interaction, respectively.  For the first term $V^\text{H}_{\text{cc}}$, we use the previously established model in the absence of depletants, with interaction strength $U_{\text{cc}}=$ 400$k_\text{B}T$. 
To estimate the depletant-depletant term we rely on additional static light scattering measurements for the small microgels (Supplementary Fig. \ref{fig:S0}, Supplementary Note \ref{fig:S0}), which lead us to an estimated Hertzian interaction strength at contact of $U_{\text{dd}}\simeq 100 k_\text{B}T$.

Assuming additive interactions in the mixture, the cross-interaction strength between large and small microgels would be $U_{\text{cd}} = 250 k_\text{B}T$.  Since simulations of the full binary system are rather costly at the small investigated size ratios, we proceed by assuming ideal depletant-depletant interactions, which simplifies the theoretical description in terms of an (effective) one component system. This assumption is justified by the small size as well as by the very soft interactions between depletants. The interactions between large microgels can thus be calculated as $V^\text{H}_{\text{eff,cc}}=V^\text{H}_{\text{cc}} + V_{\text{depl}}$, where $V_{\text{depl}}$ is the additional depletion term induced by the small microgels which depends only on the cross-interactions  $V^\text{H}_{\text{cd}}$ and on the depletant volume fraction $\phi_{\text{eff,d}}$, as explained in the Methods.

The resulting interaction potential $V^\text{H}_{\text{eff,cc}}$ is far too attractive (Supplementary Fig. \ref{fig:MH-pot}a, Supplementary Note \ref{fig:MH-pot}) even at very low $\phi_{\text{eff,d}}$, independently on the choice of $U_{\text{cd}}$. Indeed, even considering rather low (and strongly non-additive) depletant-colloid interactions, the resulting effective potential would lead to instantaneous aggregation between the colloids. In contrast, all studied binary mixtures are experimentally stable. Thus, the Hertzian repulsion model dramatically fails in capturing the behaviour of the particles once we add even the smallest amount of attractive depletion. \\

\textbf{The multi-Hertzian model for microgel-microgel interactions\\ } 

The Hertzian model fails to describe binary mixtures, because its soft repulsion is too weak to counteract the depletion attraction. Indeed, for soft, penetrable particles, we have to consider interactions down to $r\rightarrow 0$, where the depletion attraction can become very large\cite{Rovigatti:15}. We thus need to model the repulsion between microgels in a more realistic way, taking into account that the density profiles for individual microgels studied in this work show a core-corona structure\cite{Stieger:04, Boon:17, Gnan:17}. Hence, the addition of depletion interactions allows us to reveal the 'hidden' effect of the microgel core even without directly probing too dense regimes. 

\begin{figure}[h]
\centering
\includegraphics[width=0.9\columnwidth]{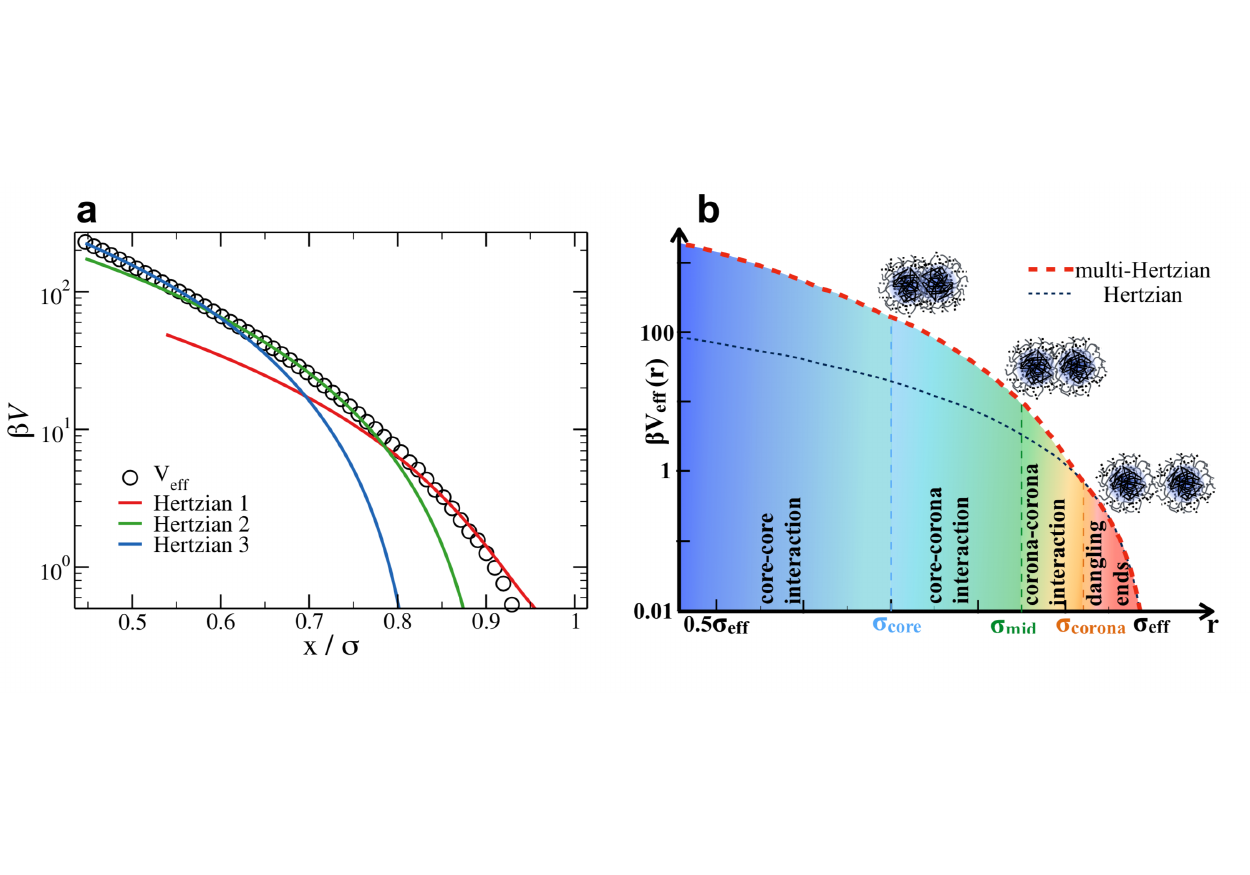} 
  \caption{{\bf The multi-Hertzian model.} (a) Calculated effective potential between two microgels as a function of their center-to-center distance. Lines are fits to three different Hertzian contributions, labelled respectively as Hertzian 1, which corresponds to the calculated elastic moduli\protect\cite{Rovigatti:18}, Hertzian 2 and Hertzian 3 representing the contributions of the inner structure of the microgels. For the reported microgel, the fitted strengths are  $U_1 = 335 k_\text{B}T$, $ U_2 = 1182 k_\text{B}T$ and $U_3 = 2617 k_\text{B}T$ and the fitted lengths are $\sigma_1 = 1.0$,  $\sigma_2 = 0.92$ and $\sigma_3= 0.8354$ in good qualitative agreement with the ones used to fit experimental data whose parameters are given in Table \protect\ref{tab:MH}; (b) the model describing experimental data with the employed interactions lengths:   
$\sigma_{\text{core}}$, below which core-core interactions take place, $\sigma_{\text{mid}}$ relevant to the onset of core-corona interactions and $\sigma_{\text{corona}}$ which reflects the heterogeneous nature of the outer corona shell.  A comparison with the Hertzian model is also provided. Note the logarithmic scales on the $y$-axis for both panels.
}
  \label{fig:schematic}
\end{figure}

In a recent numerical work, some of us have addressed the question of the validity of the Hertzian model by performing numerical simulations of realistic {\it in silico} microgel particles\cite{Gnan:17,Rovigatti:18}. We have shown that the Hertzian predictions only hold up to repulsion strengths of $\approx 6k_\text{B}T$ and to packing fractions of order unity. These results confirm that, for one-component microgels in the range of $\phi$ investigated here, we can successfully describe the system properties with the Hertzian model, with the strength of the repulsion being linked to the elastic moduli of the microgels, which can be computed independently~\cite{Rovigatti:18}.
For smaller separations, when the repulsion between two microgels sensibly exceeds the thermal energy, the interaction acquires a clear non-Hertzian nature, as shown in Figure~\ref{fig:schematic})(a). Interestingly, here we find that the full dependence of the effective interaction on the microgel-microgel separation can be fitted to a cascade (three in the example shown in Figure~\ref{fig:schematic}) of Hertzian potentials. The very good quality of the fit can be understood in terms of the microscopic architecture of the microgel, which can be considered to be composed by a sequence of more and more dense shells, each of them corresponding to a different internal elasticity and thus to a different Hertzian contribution.
Thus, while the Hertzian model is only able to capture the interactions between the outer parts of the coronas of the two microgels, stronger repulsions need to be considered in order to include core-corona and core-core contributions. Such a multi-Hertzian (MH) model is able to describe the numerically calculated potential up to the smallest simulated distances, which correspond to strengths of order $200 k_\text{B} T$ for the considered microgel.

%%%%%FIGURE 5%%%%%%%

We thus apply the MH model (schematically illustrated in Fig.~\ref{fig:schematic}) to the investigated binary mixtures, fixing most of the model parameters according to experimental data as described in Methods.
It turns out that we need to take into account four successive shells, reading as
\begin{eqnarray}
V^{\text{MH}}_{\text{cc}}(r)&=& U_{\text{cc}}(1-r/\sigma_{\text{eff}})^{5/2}\theta(\sigma_{\text{eff}}-r)+U_{\text{corona}}(1-r/\sigma_{\text{corona}})^{5/2}\theta(\sigma_{\text{corona}}-r) \nonumber \\
&+& U_{\text{mid}}(1-r/\sigma_{\text{mid}})^{5/2}\theta(\sigma_{\text{mid}}-r)+U_{\text{core}}(1-r/\sigma_{\text{core}})^{5/2} \theta(\sigma_{\text{core}}-r) \label{eq:multi}
 \end{eqnarray}
where the outermost shell, extending up to $\sigma_{\text{eff}}$, coincides with the Hertzian model of strength $U_{\text{cc}}$. This ensures that, in the investigated regime, the behaviour of the one-component microgel system is the same for the MH model and for the Hertzian model (Supplementary Fig. \ref{fig:MHoverH}, Supplementary Note \ref{fig:MHoverH}). The size of the innermost shell is set by $\sigma_{\text{core}}$, which is the experimentally determined core size and indicates the onset of core-core interactions with a very large repulsion strength $U_{\text{core}}$.  Because the transition from the core to the corona is gradual, we introduce an intermediate shell at the mid-point $\sigma_{\text{mid}}$, signaling core-corona interactions. We find that the introduction of an additional elasticity within the outer shell (starting at  $\sigma_{\text{corona}}$, the midpoint of the corona) is  necessary to reproduce the experimental data to differentiate the contribution of the dangling ends\cite{Boon:17} of the order of $\sim k_\text{B}T$ from the corona one. This turns out to be slightly different from the numerical result in Fig.~\ref{fig:schematic}a, probably due to the small size of the investigated microgels and to the absence of true dangling ends in this representation.
At each of the characteristic lengths of the potential (see Fig.~\ref{fig:schematic}b), an associated interaction strength is estimated by simple arguments (see Methods), except for $U_{\text{corona}}$, which is adjusted to match the experimental data. The obtained strengths are in qualitative agreement with those resulting from the MH fit of the calculated effective potential.\\

\textbf{Developing the multi-Hertzian model at 15$\degree$C\\} 

We apply the MH model to binary mixtures of large ('colloid') and small ('depletant') microgels. The resulting effective potential is now the sum of the multi-Hertzian model for the direct colloid-colloid interactions and the depletion term, as $V^{\text{MH}}_{\text{eff,cc}} = V^{\text{MH}}_{\text{cc}} + V_{\text{depl}}$. The latter term contains the cross-interactions between the two types of microgels (colloid-depletant interactions), which for consistency should also take a multi-Hertzian form. However, we have explicitly checked that its inclusion makes no significant difference to using a simple Hertzian. Rather, it complicates the description, so that we stick to the simple Hertzian $V^\text{H}_{\text{cd}}$, whose strength $U_{\text{cd}}$ has yet to be determined. Also, we need to determine the effective depletant volume fraction $\phi_{\text{eff,d}}$.  
 
We start by considering $T=15\degree$C and simultaneously vary the free parameters $U_{\text{corona}}$, $U_{\text{cd}}$ and $\phi_{\text{eff,d}}$ until we find an optimal agreement to reproduce the measured $g(r)$. The resulting effective potential which best describes the experimental data is found for $U_{\text{corona}}= 8.25 \times U_{\text{cc}} = 3300k_\text{B}T$ and $U_{\text{cd}}=80k_\text{B}T$ (Supplementary Fig. \ref{fig:MH-pot}b, Supplementary Note \ref{fig:MH-pot}). Effective depletant volume fractions are $\phi_{\text{eff,d}} =$ 0.18,  0.26, 0.30 at 15$\degree$C. The calculated $g(r)$s are shown as lines in Figure \ref{fig:T15-30} and are found to reproduce the behavior of the measured data for all depletant and colloid volume fractions at the examined temperature. Particularly noteworthy is the development of an asymmetric  main peak of $g(r)$ at the highest studied volume fractions, which is accurately captured by the MH model.

These findings point out the strongly non-additive character of the interactions in microgel mixtures: indeed the colloid-depletant $U_{\text{cd}}$ is significantly lower than the average of the two individual interactions for colloid and depletant microgels. This is probably due to the ability of soft particles to deform or overlap with each other, differently from hard particles. A possible explanation is that the very small depletant here involved can quite freely interpenetrate within the corona of the large ones, modifying the cross-interactions.  The non-additivity is thus the key ingredient which allows us to explain the surprising stability of our soft binary mixtures\cite{Hoffman:06, Angioletti:14}. Indeed, thanks to this feature, particles are able to experience a much more moderate depletion attraction than what is observed in hard colloid-soft depletant mixtures \cite{Bayliss:11} and in additive soft ones\cite{Rovigatti:15}. Hence our soft mixtures will eventually phase separate only at much larger depletant concentrations.\\ 

\textbf{Using the multi-Hertzian model at higher temperatures\\} 

The incorporation of  the temperature dependence is a first real test to the robustness of the MH model. With increasing temperature, the interactions between the colloids in the binary mixtures change. Increasing the temperature has an effect not only on all interactions in the MH model, but also on the colloid-depletant cross interaction $U_{\text{cd}}$ and on the effective volume fractions $\phi_{\text{eff,c}}$ and $\phi_{\text{eff,d}}$. The two volume fractions are easily dealt with: the deswelling of the microgels (in both cases) automatically yields the volume fractions at higher temperatures (see Table \ref{tab:phieff}).  
For the MH model parameter estimate, we use the temperature dependence of the Hertzian term (see also Figure \ref{fig:coll_only}b) for the outermost corona. We further note that the core size is temperature independent, based on previously published experimental data\cite{Mohanty:14} and our own unpublished work. Thus, the intermediate shells in the MH model become thinner and their associated strengths are chosen as done for 15$\degree$C. A detailed description of the choice of parameters is given in the Methods section, but it is important to stress that the temperature dependence of the MH model has zero free parameters: everything is fixed based on experimental data and the parameters found for 15$\degree$C.
Thus the only parameter left to vary is the colloid-depletant cross interaction $U_{\text{cd}}$. Once a good agreement with experimental data is found, it is checked a posteriori that the estimated values are very reasonable and obey a roughly linear relation to temperature, analogously to $U_{\text{cc}}$ (Supplementary Fig. \ref{fig:Ucd-linear}, Supplementary Note \ref{fig:Ucd-linear}).

The  experimental $g(r)$s for the binary mixtures are compared with the simulated data in Figure \ref{fig:T15-30} for all investigated $T$. The final model parameters are reported in Table \ref{tab:MH}. We find that the MH model captures all the distinct features of the depletion attraction: the peak shift, its increase and asymmetry all emerge with increasing $\phi_{\text{eff,d}}$ (Figure \ref{fig:T15-30}). It is worth to stress that the agreement of the model with experiments spans 48 different state points and is based essentially on adjusting two parameters: the strength of the second corona shell $U_{\rm corona}$ (only determined at 15$\degree$C) for the MH model and the cross-interaction strength $U_{\text{cd}}$ (adjusted at each temperature) for the depletion interaction. Thus the present findings represent a strong test in favor of the validity of the present model.

\begin{figure}[h]
\centering
  \includegraphics{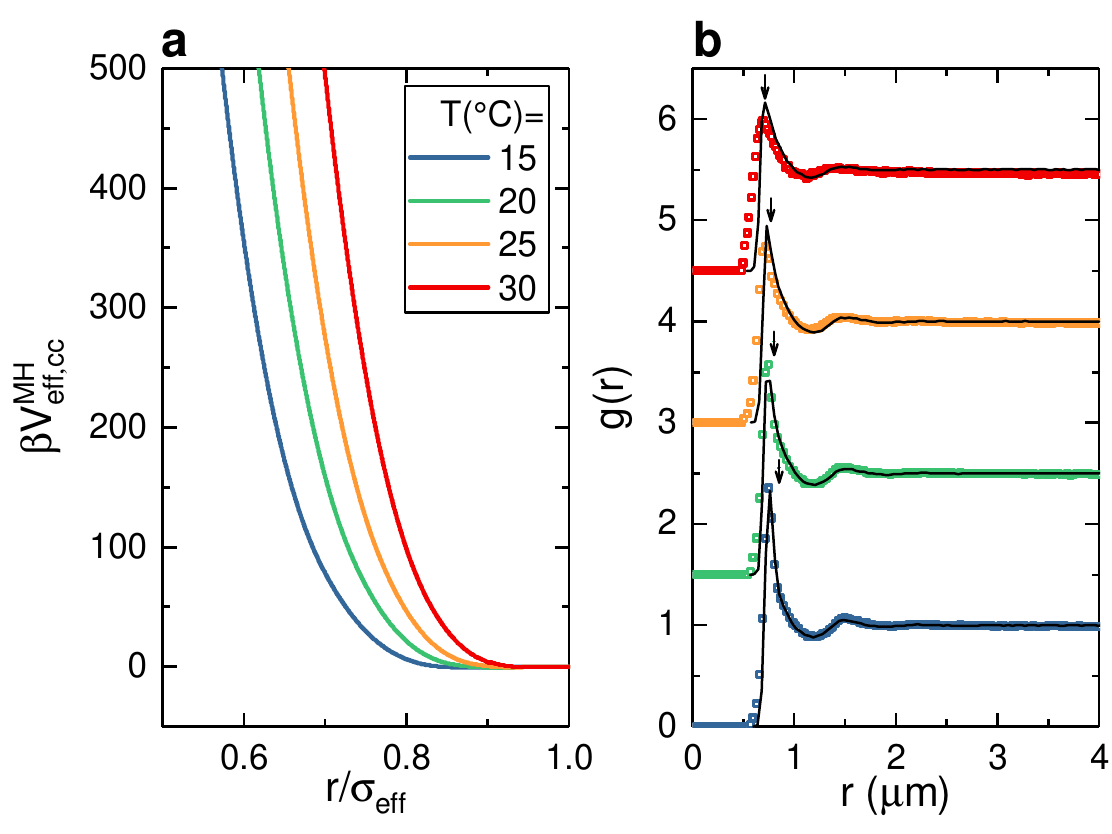}
  \caption{{\bf Typical $V^{\text{MH}}_{\text{eff,cc}}$ and resultant g(r)s for one state point across temperature.} 
  {\bf a)} Calculated effective potential $\beta V^{\text{MH}}_{\text{eff,cc}}$ based on temperature dependent $V^{\text{MH}}_{\textrm{cc}}$ and $V_{\text{cd}}$. At $T=15\degree$C, $\beta V^{\text{MH}}_{\text{eff,cc}}$ displays a shallow negative minimum at $\sim -0.6k_\text{B}T$. 
  {\bf b)} Comparison between numerical $g(r)$s (solid lines) and experimental $g(r)$s (coloured squares). From top to bottom temperature increases from 15 to 30$\degree$C. Downward pointing arrows indicate the hydrodynamic diameter of colloids.}
  \label{fig:bin_mix}
\end{figure}

In order to better visualize the effect of temperature, Figure \ref{fig:bin_mix} shows the results for the state point with largest colloid and depletant volume fractions ($\phi_{\text{eff,c}}=0.49$ and $\phi_{\text{eff,d}}=0.30$ at 15$\degree$C). An increase of $T$ again reduces the structural correlations and also the effect of depletion (due to the smaller effective depletant volume fraction), which manifests itself at each temperature by an increased asymmetry and by a shift of the main peak of the $g(r)$ toward smaller values of $r$ compared to the one-component system. The peak position is found at smaller distances with respect to the hydrodynamic radius of the colloids, clearly indicating that the particles partially overlap. The agreement of the MH model with experiments becomes worse for $30\degree$C, similarly to the case of the one-component system and probably due to the larger statistical noise in the experimental values. The average deviations between numerical and experimental curves are reported in Supplementary Fig.~\ref{fig:chi_2} and Supplementary Note \ref{fig:chi_2}.\\

%%%%FIGURE 6%%%%%

A second robustness test for the MH model is carried out by calculating the MSD and comparing it with experiments. Similarly to what has been done for the one-component systems, we rely on BD simulations and compare the calculated and measured MSD of large microgels for each of the nine mixtures in the temperature range $15\degree$C $ \leq T\leq 30\degree$C. As shown in Figure \ref{fig:msds}, the current model is also able to capture the particle dynamics for all studied state points. This is confirmed in Supplementary Fig.~\ref{fig:diff_coeff} and Supplementary Note \ref{fig:diff_coeff}, where the self-diffusion coefficients for all state points are shown and described. The small deviations between experiments and simulations observed at 30$\degree$C can again be rationalised by the larger tracking errors associated to the rapid Brownian motion of the microgels at this temperature.  \\

%------------------------------------------------

\textbf{Discussion\\}

In this study we have presented an extended investigation of microgel suspensions in a three-axis phase diagram. In addition to varying microgel volume fraction and temperature, we also varied the concentration of a second component in the suspension, namely smaller microgels, which act as depletants. We investigate one-component and binary mixtures of microgels in a wide range of control parameters, amounting to 48 different state points. Through the combination of confocal microscopy experiments and simulations we provide a systematic and comprehensive characterization of both static and dynamic observables in the form of radial distribution functions and mean-squared displacements of the large microgels. Based on explicit calculations of the effective potential between two microgels, we have been able to develop a new interaction potential that, with a single set of experiment-informed parameters, is able to reproduce the statics and dynamics of real microgel suspensions, accounting for the dependence on microgel volume fraction, temperature and depletant concentration.
Although microgels are nowadays a widely studied model system, such an extensive study was crucially missing. The several novel findings reported here will change the approach to the use of microgels as model systems in future work. Indeed, these soft particles appear to be much more complex systems than naively thought.

First of all, we have provided evidence that the effect of temperature on microgel-microgel effective interactions is not negligible, even within the swollen regime only. The soft Hertzian repulsion between the particles becomes steeper with increasing $T$. This seemingly straightforward result is not obvious since, for $T >$ VPTT, microgels become attractive due to the increased van der Waals and additional hydrophobic interactions. Therefore, an increase of repulsion goes in the opposite direction. The trend can be rationalised by thinking of the microgels only in physical terms (ignoring polymer-solvent interactions which are not yet dominant): as the particles become smaller, they also become more compact and hence somewhat less penetrable. A further change in interactions at high $T$ is however hinted by the present results, as the simple repulsive model that we have adopted shows increasing deviations at the highest studied $T=30\degree$C, approaching the VPTT at 32$\degree$C. Close to the VPTT, a much more careful evaluation, also in terms of charge effects which could become important as shown by our and others preliminary measurements\cite{truzzolillo2018overcharging}, will be required. For the examined $T$-interval, the present findings clearly show that the variation of volume fraction that is obtained by changing $T$, a commonly used method in experiments to efficiently explore a larger portion of the phase diagram, should be done with caution, as doing so significantly affects the effective interactions between the particles. Previous works have already pointed out this important aspect through indirect observations\cite{Romeo:10}, but here for the first time we provide a direct evidence and quantify the change of behavior with $T$ across the swollen regime.

Secondly, we have shown that a simple structureless model such as the Hertzian repulsion does not work to describe conditions where overlaps between particles and/or deformations start to be probed. These effects are an important physical ingredient that deeply affect the behavior of soft colloids in general and of microgels in particular, at the heart of a large research activity on glass transition and jamming of soft particles. Even without directly exploring dense conditions, the use of depletants has allowed us to probe the effective interactions between microgels at short separation distances, finding evidence of the importance of the internal microgel architecture. We have thus transferred our previous knowledge from a simple Hertzian model to a multi-Hertzian one, which is confirmed by explicit calculations of the effective potential between two microgels, that involves the inclusion of inner shells of different elasticity. Interestingly, we find that to successfully describe the experimental data it is important not only to differentiate between core and corona, but also to take into account the heterogenous character of the corona, further differentiating the contribution of the dangling ends\cite{Boon:17,Gnan:17}. Given the numerous studies where different synthesis protocols have been implemented to obtain other internal structures and crosslink density distributions, see \textit{e.g.} Refs.~\cite{acciaro:2011,tiwari:2014,wei:2016,mueller:2018}, it will be interesting to systematically study and quantify how the crosslink density and internal structure of the microgels influence their effective interactions in future studies.

The multi-Hertzian model that we have designed is based on numerical evidence and on available experimental parameters.  The comparison with experimental data has allowed us to determine the unknown parameters, most importantly the cross-interactions between small and large microgels. This is a key player in the depletion interaction, and the very low strength that we have determined for cross interactions does explain the striking finding that soft microgel mixtures are much more stable, up to very high depletant concentrations, than expected. Indeed, previous works with additive soft mixtures have shown how softness enhances depletion attraction\cite{Rovigatti:15}. Here we show that this does not happen, because softness allows deformation and interpenetration, which translates to strongly non-additive interactions. It will be interesting to  confirm these findings also for other soft mixtures and in particular,  for the more studied classical case of soft colloids and non-adsorbing polymers acting as depletants.

Finally, our phenomenological approach will have to be generalized to deal with different conditions such as even higher $T$ or larger microgel volume fractions, approaching the glass transition. However, the reported evidence clearly shows that future studies will have to explicitly take into account temperature dependence and internal microgel structure to meaningfully describe microgel behavior and to use them as model systems for exploring phase transitions and glassy dynamics.

\section*{Methods}
\subsection*{Synthesis} 
PNIPAM particles were synthesised via precipitation polymerisation \cite{pelton:86, pelton:00}. NIPAM was re-crystallised in hexane and all other chemicals were used as received.  For the large fluorescent microgels (referred to as colloids), 2.004 g N-isopropylacrylamide (NIPAM, Acros Organics) was dissolved in 82.83 g of water. 0.136 g (4.98 mol\% with respect to NIPAM) of  the cross-linker N, N-methylenebis(acrylamide) (BIS, Sigma-Aldrich) was added. 0.002 g methacryloxyethyl thiocarbonyl Rhodamine B dissolved in 10 g of water was added to the reaction mixture to covalently incorporate fluorescent sites. The reaction mixture was heated to 80$\degree$C and bubbled with nitrogen for 30 min. The reaction was then kept under a nitrogen atmosphere. To start the reaction, 0.1 g KPS in 5 g water was injected to the mixture. The reaction was then left for 4 h before the heat was turned off and the solution was left to cool down under constant stirring.
 
For the small non-fluorescent particle synthesis (referred to as depletants), we followed the same procedure. We combined 1.471 g of NIPAM, 0.0647 g (3.2 mol\% with respect to NIPAM) of BIS in 96.29 g water. 0.1929 g of sodium dodecyl sulphate (SDS, Duchefa Biochemie) was also added to induce the formation of particles with smaller radii. The mixture was heated to 70$\degree$C, bubbled with nitrogen and 0.0539 g of KPS in 2.0145 g of water was added to start the reaction. The reaction was then left for 6 hours under a nitrogen atmosphere. 

The particle suspensions were cleaned by three centrifugation and re-dispersion series before the suspensions were freeze-dried to remove all water.

\subsection*{Sample preparation}  
All samples were prepared using the freeze-dried microgels and deionised water (purified with a MilliQ system), as this allows us to control the weight concentration. In order to ensure homogeneous dispersions, samples were thoroughly mixed by vortexing and sonication followed by placing the dispersion on a tumbler for two weeks prior to any experiment. 

Using this approach, samples with a wt\%-range from 0.1 to 1wt\% of colloids and samples with wt\%-range from 0.1 to 0.8wt\% of depletants were prepared for viscometry experiments. Very dilute colloid and very dilute depletant suspensions ($<$0.1wt\%) were made for DLS characterisation. Suspensions were diluted until almost completely transparent to avoid multiple scattering. For the SLS measurements, samples with a wt\%-range from 0.05 to 0.65wt\% depletant were prepared. For the CLSM experiments, we aimed for binary mixtures with effective colloid volume fraction $\phi_{\text{eff,c}} = 0.2, 0.3, 0.4$ and with additional effective depletant volume fraction $\phi_{\text{eff,d}} = 0, 0.1, 0.2, 0.3$ at 15$\degree$C. As an initial guess for the packing fraction of the samples, we used the shift factor $k=\phi / {\text{wt\%}}$ as determined from viscometry measurements on colloid-only and depletant-only samples (see below for the experimental $k$-values). The binary mixtures contained colloid wt\% 2.2, 3.3 and 4.4 and depletant wt\% 0, 0.26, 0.54, 0.81. Final $\phi_{\text{eff}}$ were determined by fitting $g(r)$ curves, as discussed in the manuscript.

\subsection*{Experiments} 
The viscosity of colloid-only and depletant-only samples with known wt\%-concentration was recorded using an Ubbelohde viscometer at 15 and 30$\degree$C. Flow times were measured 5-6 times, averaged and divided by the flowtime of a water sample to extract the relative viscosity of the samples. The relative viscosity was fitted to the well-known Batchelor equation which holds for colloids in the dilute regime \cite{batchelor:77}: $\eta_{\text{rel}} = 1 + 2.5\phi_{\text{eff}} + 5.9\phi_{\text{eff}}^2$ with $\phi_{\text{eff}} = k \times $wt\%. From these fits the shift factor $k$ was determined for 15 and 30$\degree$C. For 20, 25$\degree$C, the data was interpolated. $k_{\text{colloid}}=0.091, 0.0758, 0.061, 0.046$ wt\%$^{-1}$, $k_{\text{depletant}}=0.332, 0.292, 0.253, 0.214$ wt\%$^{-1}$ for 15-30$\degree$C respectively. The shift factor was used in sample preparation to estimate $\phi_{\text{eff}}$. 

Microgels were characterised using dynamic light scattering (DLS) with a goniometer-based light scattering instrument that employs pseudo 2D-cross correlation (3D DLS Spectrometer, LS instruments, Switzerland) with laser wavelength $\lambda=660$ nm. DLS measurements were performed over a range of 15-30~$^\circ$C resulting in a swelling curve for both colloids and depletants. The hydrodynamic radii were extracted using a first order cumulant analysis averaged over an angular range of 60-100$^\circ$, and measured every 10$^\circ$. To probe the interactions between depletants, static light scattering experiments were performed at several packing fractions, and the small wavevector limit $S(0)$ of the static structure factor $S(q)$ was also obtained.

The binary mixtures were imaged in a Leica SP5 confocal microscope at a frame rate of 13.9 Hz in the range of 15-30$\degree$C. An excitation wavelength of 543 nm was used in combination with an oil immersion objective at 100x magnification and numerical aperture 1.4. The confocal microscope is housed in a temperature regulated box which provides a temperature control with a stability of $\pm$ 0.2$\degree$C over the range of temperatures used. Because scanning in the $z$-direction would have been too slow, we made $xyt$-videos. Such videos of 512x512x4000 frames were obtained for at least five different positions in the sample to minimise the effects of local density fluctuations. Videos were taken at $\gtrsim$ 5 particle diameters away from the glass to avoid wall influences. The accuracy of the coordinates is estimated to be $\Delta x \approx \Delta y \approx 11$nm \cite{Mohanty:14}.  Using standardised image analysis and particle tracking routines \cite{Crocker:96}, the 2D $g(r)$s and 2D mean square displacements (MSDs) ($\langle x^2+y^2\rangle$) were obtained. To ensure the 2D $g(r)$ corresponds to the 3D $g(r)$, the approach as described in Mohanty et al. was employed \cite{Mohanty:14}. In brief, a thinner 'slice' of data is created by rejecting out of focus particles, i.e. we only take particles with $z=0$. Even so, there will always remain some variation in the $z$-position of the tracked particles. This has been taken into account in the numerical calculations by adding a suitable noise along one of the axes.

\subsection*{Model and theory} 
We consider two systems: one-component microgel systems and binary mixtures. Colloids experience a direct colloid-colloid interaction that we model as Hertzian or multi-Hertzian (MH) as described in the manuscript. The presence of depletants leads to an additional attractive interaction between the colloids. Thus, the effective colloid-colloid interaction potential $V_{\text{eff,cc}}$ is the sum of two contributions: $V_{\text{cc}}$ (direct colloid-colloid interaction) and $V_{\text{depl}}$ (depletion interaction), i.e. $V_{\text{eff,cc}}(r) = V_{\text{cc}}(r) + V_{\text{depl}}(r)$. 
We assume that depletants are ideal. Under this assumption, the Fourier components of the additional depletion term can be calculated for any colloid-depletant interaction $V_{\text{cd}}$ as in Ref.\cite{Parola:15}:
\begin{equation}
-\beta \tilde{V}_{\text{depl}}(k)= \rho_d\left[\int d{\bf r} (e^{-\beta V_{\text{cd}}(r)}-1) e^{i {\bf k}\cdot{\bf r}}\right]^2
\label{eq:depl}
\end{equation}
Here $\rho_\text{d}$ is the reservoir depletant number density and $\beta=1/(k_\text{B}T)$. In our study, we consider $V_{\text{cd}}(r)$ to be a Hertzian potential. After Fourier transforming Eq.~\ref{eq:depl} we obtain $\beta V_{\text{depl}} (r)$ which is added to $\beta V_{\text{cc}}$ at each considered temperature for the binary mixtures to obtain the total interaction potential $\beta V_{\text{eff,cc}}$. We further check that the use of a multi-Hertzian model for $V_{\text{cd}}(r)$ does not yield a noticeable change on the obtained results.

The MH model is built up as follows. The outer shell corresponds to the Hertzian soft repulsion: $U_{\text{cc}}=400, 520, 640, 760k_\text{B}T$ for 15, 20, 25, 30$\degree$C which sets in at $r=\sigma_{\text{eff}}$, where $\sigma_{\text{eff}} = 2R_\text{H}$ is of course temperature dependent (see also Supplementary Fig. \ref{fig:swelling} and Supplementary Note \ref{fig:swelling}). The inner shell corresponds to the core and is temperature independent. We estimate the core diameter as $\sigma_{\text{core}}=0.7\sigma_{\text{eff}}$ thanks to available experimental SAXS data (and related fuzzy-sphere model fits) for similar microgels \cite{Mohanty:14, jeromemanuscript}. We fix $U_{\text{core}}=10^4k_\text{B}T$, a value compatible with elasticity arguments \cite{Vlassopoulos:14} which takes into account the high crosslink density in the core. In this way, the innermost and outermost Hertzian terms are completely specified based on experimental data.  Since the border between the dense core and loosely crosslinked corona is not so well-defined, we introduce an intermediate point at the mid-point between these two lengths, i.e. $\sigma_{\text{mid}} = 0.5(\sigma_{\text{core}} + \sigma_{\text{eff}}) = 0.85\sigma_{\text{eff}}$.  The associated repulsion strength is arbitrarily chosen to be $U_{\text{mid}} = 10 \times U_{\text{cc}}$. We find that it is crucial to describe the corona by using two distinct shells, introducing a second intermediate point $\sigma_{\text{corona}}=0.5(\sigma_{\text{mid}}+\sigma_{\text{eff}})= 0.925\sigma_{\text{eff}}$ with its associated strength $U_{\text{corona}}$. The latter is determined at 15$\degree$C to be $8.25 \times U_{\text{cc}}$ by comparing simulation results with the experimental $g(r)$s. This relation and that for $U_{\text{mid}} = 10 \times U_{\text{cc}}$ are then kept throughout the temperature range.

The effective potential $V_{\text{eff,cc}}$ is then calculated at several depletant volume fractions starting with our initial guess from viscometry. Because of the uncertainty in the experimental packing fraction, we adjust $\phi_{\text{eff,d}}$ in the simulations at 15$\degree$C until we find good agreement with the experimental data. As described in the manuscript, $\phi_{\text{eff,c}}$ was adjusted in the Hertzian simulations at 15$\degree$C. The thus obtained parameter values for the volume fractions ($\phi_{\text{eff,c}}, \phi_{\text{eff,d}}$) and the parameters for the MH model are summarised in Tables \ref{tab:phieff} and \ref{tab:MH}, respectively.

%%%%TABLE 1%%%%%

%%%%TABLE 2%%%%%

To justify the assumption of ideal behavior of small microgels  and the non-additive character of interactions in the mixture, we have further quantified the interactions between small microgels assuming a Hertzian repulsion. To calibrate its strength, we have computed $S(0)$ by solving the Ornstein-Zernike equation within the Rogers-Young closure, finding that a very soft interaction between depletants, i.e. $U_{\text{dd}} \simeq 100k_\text{B}T$, captures the small microgels behavior. This estimate is consistent with scaling arguments of the Hertzian model as a function of particle size\cite{landau:58,Riest:12} with respect to the large microgels. 

\subsection*{Numerical calculation of the effective potential} 
Microgel configurations are built as disordered, fully-bonded networks generated as in Ref.~\cite{Gnan:17,Rovigatti:18} using $\approx 5000$ monomers of diameter $\sigma_m$ in a spherical confinement of radius $Z = 25$~$\sigma_m$ and crosslinker concentration $c=5\%$. Monomers are in the swollen regime and interact through the classical bead-spring model for polymers~\cite{grest1986molecular}. 

To calculate the microgel-microgel effective interactions we combine two methods, as described in Ref.~\cite{Rovigatti:18}. We perform umbrella sampling simulations in which we add an harmonic biasing potential acting on the centres of mass of the two microgels~\cite{roux1995calculation} for small separation distances $r$. The resulting effective potential is calculated as $V(r)= -k_\text{B} T \log[g(r)]$. At larger values of $r$, we employ a generalised Widom insertion scheme~\cite{mladek2011pair} which is very efficient in sampling the small-deformation regime.

\subsection*{Bulk simulations} 
We perform Langevin dynamics simulations of $N = 2000$ colloid particles of mass $m$ interacting with the generated effective interaction potential $V_{\text{eff,cc}}$. The units of length, energy and mass are $\sigma_{\text{eff}}$, $k_\text{B} T$ and $m$ respectively. Time is measured in units of $\sqrt{m\sigma_{\text{eff}}^2/k_\text{B}T}$. The integration time-step is fixed to 0.001. With this scheme, particles after an initial microscopic time follow a Brownian Dynamics (BD) due to the interactions with a fictitious solvent\cite{Allen}. The solvent effective viscosity enters in the definition of the zero-colloid limit self-diffusion coefficient $D_0$, which is the key parameter in BD simulations.  

Since the viscosity of the real samples changes for each $T$ and for each $\phi_{\text{eff,d}}$, we have estimated the experimental $D_0$ at $\phi_{\text{eff,d}}=0$ by means of Stokes-Einstein relations using the measured hydrodynamic radii at each $T$. Furthermore, we have also estimated $D_0$ in the presence of depletant thanks to the viscosity measurements described above.  Ideally, one would need to directly use these values in the simulations, but this leads to an incorrect description of the system at low enough colloid packing fractions because BD simulations do not include hydrodynamic interactions, whose effects are strongest in this regime. Hence we have adopted the following strategy: at fixed $\phi_{\text{eff,c}}$ and for each $T$ and $\phi_{\text{eff,d}}$ (i.e. for 16 of our samples), we performed several BD runs in order to select the values of $D_0$ providing a good agreement for the long-time MSD with experiments. We thus find a unique shift factor on the time axis for all simulated state points, needed in order to convert simulation time into experimental time.
Then, the estimated $D_0$ values were kept fixed for all studied $\phi_{\text{eff,c}}$, i.e. for the remaining 32 studied samples no further adjustment was made.
The estimated $D_0$ can thus be considered the effective bare self-diffusion coefficients of our approximated BD approach, and was finally compared to the experimental estimates, finding a good agreement at large depletant concentrations and low temperatures (as reported in Supplementary Fig.~\ref{fig:D0} and Supplementary note \ref{fig:D0}), that is, for the state points where hydrodynamic interactions are less important.

Simulations were performed with particles possessing a polydispersity of 4\% with a Gaussian distribution, similar to the experimental system. Slices through configurations of 100 independent state points were used to calculate the radial distribution function $g(r)$ of the 3D data with sufficient statistics. The $z$-position of particles is randomly displaced by Gaussian noise with a standard deviation of 0.005. In this way, $g(r)$s can be successfully compared to the 2D-$g(r)$ obtained from experiments, as demonstrated in Ref.\cite{Mohanty:14}. 

\subsection*{Data availability}
The authors declare that all data supporting the findings of this study are available within the article and its Supplementary Information files. All other relevant data supporting the findings of this study are available from the corresponding authors on request.

\subsection*{Code availability}
The computer codes used for the current study are available from the corresponding authors on reasonable request.
\clearpage
%----------------------------------------------------------------------------------------
%	REFERENCE LIST
%----------------------------------------------------------------------------------------
%

\bibliography{references-use} 

\begin{thebibliography}{10}
\expandafter\ifx\csname url\endcsname\relax
  \def\url#1{\texttt{#1}}\fi
\expandafter\ifx\csname urlprefix\endcsname\relax\def\urlprefix{URL }\fi
\providecommand{\bibinfo}[2]{#2}
\providecommand{\eprint}[2][]{\url{#2}}

\bibitem{Vlassopoulos:14}
\bibinfo{author}{Vlassopoulos, D.} \& \bibinfo{author}{Cloitre, M.}
\newblock \bibinfo{title}{Tunable rheology of dense soft deformable colloids}.
\newblock \emph{\bibinfo{journal}{Curr. Opin. Colloid In.}}
  \textbf{\bibinfo{volume}{19}}, \bibinfo{pages}{561--574}
  (\bibinfo{year}{2014}).

\bibitem{Vanderscheer:17}
\bibinfo{author}{van~der Scheer, P.}, \bibinfo{author}{van~de Laar, T.},
  \bibinfo{author}{van~der Gucht, J.}, \bibinfo{author}{Vlassopoulos, D.} \&
  \bibinfo{author}{Sprakel, J.}
\newblock \bibinfo{title}{Fragility and strength in nanoparticle glasses}.
\newblock \emph{\bibinfo{journal}{ACS Nano}} \textbf{\bibinfo{volume}{11}},
  \bibinfo{pages}{6755--6763} (\bibinfo{year}{2017}).

\bibitem{Gaurasundar2017}
\bibinfo{author}{Conley, G.~M.}, \bibinfo{author}{Aebischer, P.},
  \bibinfo{author}{N{\"o}jd, S.}, \bibinfo{author}{Schurtenberger, P.} \&
  \bibinfo{author}{Scheffold, F.}
\newblock \bibinfo{title}{Jamming and overpacking fuzzy microgels: Deformation,
  interpenetration, and compression}.
\newblock \emph{\bibinfo{journal}{Sci. Adv.}} \textbf{\bibinfo{volume}{3}},
  \bibinfo{pages}{e1700969} (\bibinfo{year}{2017}).

\bibitem{Mohanty2017}
\bibinfo{author}{Mohanty, P.~S.} \emph{et~al.}
\newblock \bibinfo{title}{Interpenetration of polymeric microgels at ultrahigh
  densities.}
\newblock \emph{\bibinfo{journal}{Sci. Rep.}} \textbf{\bibinfo{volume}{7}},
  \bibinfo{pages}{1487} (\bibinfo{year}{2017}).

\bibitem{Stieger:04}
\bibinfo{author}{Stieger, M.}, \bibinfo{author}{Pedersen, J.~S.},
  \bibinfo{author}{Lindner, P.} \& \bibinfo{author}{Richtering, W.}
\newblock \bibinfo{title}{Are thermoresponsive microgels model systems for
  concentrated colloidal suspensions? a rheology and small-angle neutron
  scattering study}.
\newblock \emph{\bibinfo{journal}{Langmuir}} \textbf{\bibinfo{volume}{20}},
  \bibinfo{pages}{7283--7292} (\bibinfo{year}{2004}).

\bibitem{Rey:16}
\bibinfo{author}{Rey, M.} \emph{et~al.}
\newblock \bibinfo{title}{Isostructural solid--solid phase transition in
  monolayers of soft core--shell particles at fluid interfaces: structure and
  mechanics}.
\newblock \emph{\bibinfo{journal}{Soft Matter}} \textbf{\bibinfo{volume}{12}},
  \bibinfo{pages}{3545--3557} (\bibinfo{year}{2016}).

\bibitem{Boon:17}
\bibinfo{author}{Boon, N.} \& \bibinfo{author}{Schurtenberger, P.}
\newblock \bibinfo{title}{Swelling of micro-hydrogels with a crosslinker
  gradient.}
\newblock \emph{\bibinfo{journal}{Phys. Chem. Chem. Phys.}}
  \textbf{\bibinfo{volume}{19}}, \bibinfo{pages}{23740--23746}
  (\bibinfo{year}{2017}).

\bibitem{Gnan:17}
\bibinfo{author}{Gnan, N.}, \bibinfo{author}{Rovigatti, L.},
  \bibinfo{author}{Bergman, M.} \& \bibinfo{author}{Zaccarelli, E.}
\newblock \bibinfo{title}{In silico synthesis of microgel particles}.
\newblock \emph{\bibinfo{journal}{Macromolecules}}
  \textbf{\bibinfo{volume}{50}}, \bibinfo{pages}{8777--8786}
  (\bibinfo{year}{2017}).

\bibitem{mohanty:16}
\bibinfo{author}{Mohanty, P.} \emph{et~al.}
\newblock \bibinfo{title}{Dielectric spectroscopy of ionic microgel
  suspensions}.
\newblock \emph{\bibinfo{journal}{Soft Matter}} \textbf{\bibinfo{volume}{12}},
  \bibinfo{pages}{9705--9727} (\bibinfo{year}{2016}).

\bibitem{hashmi2009mechanical}
\bibinfo{author}{Hashmi, S.~M.} \& \bibinfo{author}{Dufresne, E.~R.}
\newblock \bibinfo{title}{Mechanical properties of individual microgel
  particles through the deswelling transition}.
\newblock \emph{\bibinfo{journal}{Soft Matter}} \textbf{\bibinfo{volume}{5}},
  \bibinfo{pages}{3682--3688} (\bibinfo{year}{2009}).

\bibitem{bachman2015ultrasoft}
\bibinfo{author}{Bachman, H.} \emph{et~al.}
\newblock \bibinfo{title}{Ultrasoft, highly deformable microgels}.
\newblock \emph{\bibinfo{journal}{Soft Matter}} \textbf{\bibinfo{volume}{11}},
  \bibinfo{pages}{2018--2028} (\bibinfo{year}{2015}).

\bibitem{Reese:04}
\bibinfo{author}{Reese, C.~E.}, \bibinfo{author}{Mikhonin, A.~V.},
  \bibinfo{author}{Kamenjicki, M.}, \bibinfo{author}{Tikhonov, A.} \&
  \bibinfo{author}{Asher, S.~A.}
\newblock \bibinfo{title}{Nanogel nanosecond photonic crystal optical
  switching}.
\newblock \emph{\bibinfo{journal}{J. Am. Chem. Soc.}}
  \textbf{\bibinfo{volume}{126}}, \bibinfo{pages}{1493--1496}
  (\bibinfo{year}{2004}).

\bibitem{Serpe2012}
\bibinfo{author}{Serpe, M.~J.} \emph{et~al.}
\newblock \emph{\bibinfo{title}{Hydrogel Micro and Nanoparticles}}, chap.
  \bibinfo{chapter}{Color-Tunable Poly (N-Isopropylacrylamide) Microgel-Based
  Etalons: Fabrication, Characterization, and Applications},
  \bibinfo{pages}{317--336} (\bibinfo{publisher}{Wiley-VCH Verlag GmbH and Co.
  KGaA}, \bibinfo{year}{2012}).

\bibitem{Hamidi:08}
\bibinfo{author}{Hamidi, M.}, \bibinfo{author}{Azadi, A.} \&
  \bibinfo{author}{Rafiei, P.}
\newblock \bibinfo{title}{Hydrogel nanoparticles in drug delivery}.
\newblock \emph{\bibinfo{journal}{Adv. Drug Deliver. Rev.}}
  \textbf{\bibinfo{volume}{60}}, \bibinfo{pages}{1638--1649}
  (\bibinfo{year}{2008}).

\bibitem{Peppas:00}
\bibinfo{author}{Peppas, N.}, \bibinfo{author}{Bures, P.},
  \bibinfo{author}{Leobandung, W.} \& \bibinfo{author}{Ichikawa, H.}
\newblock \bibinfo{title}{Hydrogels in pharmaceutical formulations}.
\newblock \emph{\bibinfo{journal}{Eur. J. Pharm. Biopharm.}}
  \textbf{\bibinfo{volume}{50}}, \bibinfo{pages}{27--46}
  (\bibinfo{year}{2000}).

\bibitem{oh:08}
\bibinfo{author}{Oh, J.~K.}, \bibinfo{author}{Drumright, R.},
  \bibinfo{author}{Siegwart, D.~J.} \& \bibinfo{author}{Matyjaszewski, K.}
\newblock \bibinfo{title}{The development of microgels/nanogels for drug
  delivery applications}.
\newblock \emph{\bibinfo{journal}{Prog. Polym. Sci.}}
  \textbf{\bibinfo{volume}{33}}, \bibinfo{pages}{448--477}
  (\bibinfo{year}{2008}).

\bibitem{fernandez:09}
\bibinfo{author}{Fern{\'a}ndez-Barbero, A.} \emph{et~al.}
\newblock \bibinfo{title}{Gels and microgels for nanotechnological
  applications}.
\newblock \emph{\bibinfo{journal}{Adv. Colloid Interfac.}}
  \textbf{\bibinfo{volume}{147}}, \bibinfo{pages}{88--108}
  (\bibinfo{year}{2009}).

\bibitem{wang:12}
\bibinfo{author}{Wang, Z.}, \bibinfo{author}{Wang, F.}, \bibinfo{author}{Peng,
  Y.}, \bibinfo{author}{Zheng, Z.} \& \bibinfo{author}{Han, Y.}
\newblock \bibinfo{title}{Imaging the homogeneous nucleation during the melting
  of superheated colloidal crystals}.
\newblock \emph{\bibinfo{journal}{Science}} \textbf{\bibinfo{volume}{338}},
  \bibinfo{pages}{87--90} (\bibinfo{year}{2012}).

\bibitem{hilhorst:11}
\bibinfo{author}{Hilhorst, J.} \& \bibinfo{author}{Petukhov, A.}
\newblock \bibinfo{title}{Variable dislocation widths in colloidal crystals of
  soft thermosensitive spheres}.
\newblock \emph{\bibinfo{journal}{Phys. Rev. Lett.}}
  \textbf{\bibinfo{volume}{107}}, \bibinfo{pages}{095501}
  (\bibinfo{year}{2011}).

\bibitem{peng:10}
\bibinfo{author}{Peng, Y.}, \bibinfo{author}{Wang, Z.},
  \bibinfo{author}{Alsayed, A.~M.}, \bibinfo{author}{Yodh, A.~G.} \&
  \bibinfo{author}{Han, Y.}
\newblock \bibinfo{title}{Melting of colloidal crystal films}.
\newblock \emph{\bibinfo{journal}{Phys. Rev. Lett.}}
  \textbf{\bibinfo{volume}{104}}, \bibinfo{pages}{205703}
  (\bibinfo{year}{2010}).

\bibitem{alsayed:05}
\bibinfo{author}{Alsayed, A.~M.}, \bibinfo{author}{Islam, M.~F.},
  \bibinfo{author}{Zhang, J.}, \bibinfo{author}{Collings, P.~J.} \&
  \bibinfo{author}{Yodh, A.~G.}
\newblock \bibinfo{title}{Premelting at defects within bulk colloidal
  crystals}.
\newblock \emph{\bibinfo{journal}{Science}} \textbf{\bibinfo{volume}{309}},
  \bibinfo{pages}{1207--1210} (\bibinfo{year}{2005}).

\bibitem{Mohanty:15}
\bibinfo{author}{Mohanty, P.~S.}, \bibinfo{author}{Bagheri, P.},
  \bibinfo{author}{N{\"o}jd, S.}, \bibinfo{author}{Yethiraj, A.} \&
  \bibinfo{author}{Schurtenberger, P.}
\newblock \bibinfo{title}{Multiple path-dependent routes for phase-transition
  kinetics in thermoresponsive and field-responsive ultrasoft colloids}.
\newblock \emph{\bibinfo{journal}{Phys. Rev. X}} \textbf{\bibinfo{volume}{5}},
  \bibinfo{pages}{011030} (\bibinfo{year}{2015}).

\bibitem{Zhang:09}
\bibinfo{author}{Zhang, Z.} \emph{et~al.}
\newblock \bibinfo{title}{Thermal vestige of the zero-temperature jamming
  transition}.
\newblock \emph{\bibinfo{journal}{Nature}} \textbf{\bibinfo{volume}{459}},
  \bibinfo{pages}{230--233} (\bibinfo{year}{2009}).

\bibitem{Caswell:13}
\bibinfo{author}{Caswell, T.~A.}, \bibinfo{author}{Zhang, Z.},
  \bibinfo{author}{Gardel, M.~L.} \& \bibinfo{author}{Nagel, S.~R.}
\newblock \bibinfo{title}{Observation and characterization of the vestige of
  the jamming transition in a thermal three-dimensional system}.
\newblock \emph{\bibinfo{journal}{Phys. Rev. E}} \textbf{\bibinfo{volume}{87}},
  \bibinfo{pages}{012303} (\bibinfo{year}{2013}).

\bibitem{Yunker:14}
\bibinfo{author}{Yunker, P.~J.} \emph{et~al.}
\newblock \bibinfo{title}{Physics in ordered and disordered colloidal matter
  composed of poly (n-isopropylacrylamide) microgel particles}.
\newblock \emph{\bibinfo{journal}{Rep. Prog. Phys.}}
  \textbf{\bibinfo{volume}{77}}, \bibinfo{pages}{056601}
  (\bibinfo{year}{2014}).

\bibitem{Heskins:68}
\bibinfo{author}{Heskins, M.} \& \bibinfo{author}{Guillet, J.~E.}
\newblock \bibinfo{title}{Solution properties of poly (n-isopropylacrylamide)}.
\newblock \emph{\bibinfo{journal}{J. Macromol Sci. Chem.}}
  \textbf{\bibinfo{volume}{2}}, \bibinfo{pages}{1441--1455}
  (\bibinfo{year}{1968}).

\bibitem{pelton:00}
\bibinfo{author}{Pelton, R.}
\newblock \bibinfo{title}{Temperature-sensitive aqueous microgels}.
\newblock \emph{\bibinfo{journal}{Adv. Colloid Interfac.}}
  \textbf{\bibinfo{volume}{85}}, \bibinfo{pages}{1--33} (\bibinfo{year}{2000}).

\bibitem{pelton:86}
\bibinfo{author}{Pelton, R.} \& \bibinfo{author}{Chibante, P.}
\newblock \bibinfo{title}{Preparation of aqueous latices with
  n-isopropylacrylamide}.
\newblock \emph{\bibinfo{journal}{Colloid. Surface}}
  \textbf{\bibinfo{volume}{20}}, \bibinfo{pages}{247--256}
  (\bibinfo{year}{1986}).

\bibitem{Romeo:10}
\bibinfo{author}{Romeo, G.}, \bibinfo{author}{Fernandez-Nieves, A.},
  \bibinfo{author}{Wyss, H.~M.}, \bibinfo{author}{Acierno, D.} \&
  \bibinfo{author}{Weitz, D.~A.}
\newblock \bibinfo{title}{Temperature-controlled transitions between glass,
  liquid, and gel states in dense p-nipa suspensions}.
\newblock \emph{\bibinfo{journal}{Adv. Mater.}} \textbf{\bibinfo{volume}{22}},
  \bibinfo{pages}{3441--3445} (\bibinfo{year}{2010}).

\bibitem{Heyes:09}
\bibinfo{author}{Heyes, D.} \& \bibinfo{author}{Bra{\'n}ka, A.}
\newblock \bibinfo{title}{Interactions between microgel particles}.
\newblock \emph{\bibinfo{journal}{Soft Matter}} \textbf{\bibinfo{volume}{5}},
  \bibinfo{pages}{2681--2685} (\bibinfo{year}{2009}).

\bibitem{Senff:99}
\bibinfo{author}{Senff, H.} \& \bibinfo{author}{Richtering, W.}
\newblock \bibinfo{title}{Temperature sensitive microgel suspensions: Colloidal
  phase behavior and rheology of soft spheres}.
\newblock \emph{\bibinfo{journal}{J. Chem. Phys.}}
  \textbf{\bibinfo{volume}{111}}, \bibinfo{pages}{1705--1711}
  (\bibinfo{year}{1999}).

\bibitem{Wu:031}
\bibinfo{author}{Wu, J.}, \bibinfo{author}{Huang, G.} \& \bibinfo{author}{Hu,
  Z.}
\newblock \bibinfo{title}{Interparticle potential and the phase behavior of
  temperature-sensitive microgel dispersions}.
\newblock \emph{\bibinfo{journal}{Macromolecules}}
  \textbf{\bibinfo{volume}{36}}, \bibinfo{pages}{440--448}
  (\bibinfo{year}{2003}).

\bibitem{Mohanty:14}
\bibinfo{author}{Mohanty, P.~S.}, \bibinfo{author}{Paloli, D.},
  \bibinfo{author}{Crassous, J.~J.}, \bibinfo{author}{Zaccarelli, E.} \&
  \bibinfo{author}{Schurtenberger, P.}
\newblock \bibinfo{title}{Effective interactions between soft-repulsive
  colloids: Experiments, theory, and simulations}.
\newblock \emph{\bibinfo{journal}{J. Chem. Phys.}}
  \textbf{\bibinfo{volume}{140}}, \bibinfo{pages}{094901}
  (\bibinfo{year}{2014}).

\bibitem{Scheffold:10}
\bibinfo{author}{Scheffold, F.} \emph{et~al.}
\newblock \bibinfo{title}{Brushlike interactions between thermoresponsive
  microgel particles}.
\newblock \emph{\bibinfo{journal}{Phys. Rev. Lett.}}
  \textbf{\bibinfo{volume}{104}}, \bibinfo{pages}{128304}
  (\bibinfo{year}{2010}).

\bibitem{Romeo:13}
\bibinfo{author}{Romeo, G.} \& \bibinfo{author}{Ciamarra, M.~P.}
\newblock \bibinfo{title}{Elasticity of compressed microgel suspensions}.
\newblock \emph{\bibinfo{journal}{Soft Matter}} \textbf{\bibinfo{volume}{9}},
  \bibinfo{pages}{5401--5406} (\bibinfo{year}{2013}).

\bibitem{Paloli:12}
\bibinfo{author}{Paloli, D.}, \bibinfo{author}{Mohanty, P.~S.},
  \bibinfo{author}{Crassous, J.~J.}, \bibinfo{author}{Zaccarelli, E.} \&
  \bibinfo{author}{Schurtenberger, P.}
\newblock \bibinfo{title}{Fluid--solid transitions in soft-repulsive colloids}.
\newblock \emph{\bibinfo{journal}{Soft Matter}} \textbf{\bibinfo{volume}{9}},
  \bibinfo{pages}{3000--3004} (\bibinfo{year}{2013}).

\bibitem{Likos:01}
\bibinfo{author}{Likos, C.~N.}
\newblock \bibinfo{title}{Effective interactions in soft condensed matter
  physics}.
\newblock \emph{\bibinfo{journal}{Phys. Rep.}} \textbf{\bibinfo{volume}{348}},
  \bibinfo{pages}{267--439} (\bibinfo{year}{2001}).

\bibitem{Bayliss:11}
\bibinfo{author}{Bayliss, K.}, \bibinfo{author}{Van~Duijneveldt, J.},
  \bibinfo{author}{Faers, M.} \& \bibinfo{author}{Vermeer, A.}
\newblock \bibinfo{title}{Comparing colloidal phase separation induced by
  linear polymer and by microgel particles}.
\newblock \emph{\bibinfo{journal}{Soft Matter}} \textbf{\bibinfo{volume}{7}},
  \bibinfo{pages}{10345--10352} (\bibinfo{year}{2011}).

\bibitem{Rovigatti:15}
\bibinfo{author}{Rovigatti, L.}, \bibinfo{author}{Gnan, N.},
  \bibinfo{author}{Parola, A.} \& \bibinfo{author}{Zaccarelli, E.}
\newblock \bibinfo{title}{How soft repulsion enhances the depletion mechanism}.
\newblock \emph{\bibinfo{journal}{Soft Matter}} \textbf{\bibinfo{volume}{11}},
  \bibinfo{pages}{692--700} (\bibinfo{year}{2015}).

\bibitem{Rovigatti:18}
\bibinfo{author}{Rovigatti, L.}, \bibinfo{author}{Gnan, N.},
  \bibinfo{author}{Rovigatti, L.}, \bibinfo{author}{Ninarello, A.} \&
  \bibinfo{author}{Zaccarelli, E.}
\newblock \bibinfo{title}{On the validity of the hertzian model: the case of
  soft colloids}.
\newblock \emph{\bibinfo{journal}{Preprint at arXiv:1808.04769}}
  (\bibinfo{year}{2018}).

\bibitem{Hoffman:06}
\bibinfo{author}{Hoffmann, N.}, \bibinfo{author}{Ebert, F.},
  \bibinfo{author}{Likos, C.~N.}, \bibinfo{author}{L{\"o}wen, H.} \&
  \bibinfo{author}{Maret, G.}
\newblock \bibinfo{title}{Partial clustering in binary two-dimensional
  colloidal suspensions}.
\newblock \emph{\bibinfo{journal}{Phys. Rev. Lett.}}
  \textbf{\bibinfo{volume}{97}}, \bibinfo{pages}{078301}
  (\bibinfo{year}{2006}).

\bibitem{Angioletti:14}
\bibinfo{author}{Angioletti-Uberti, S.}, \bibinfo{author}{Varilly, P.},
  \bibinfo{author}{Mognetti, B.~M.} \& \bibinfo{author}{Frenkel, D.}
\newblock \bibinfo{title}{Mobile linkers on dna-coated colloids: valency
  without patches}.
\newblock \emph{\bibinfo{journal}{Phys. Rev. Lett.}}
  \textbf{\bibinfo{volume}{113}}, \bibinfo{pages}{128303}
  (\bibinfo{year}{2014}).

\bibitem{truzzolillo2018overcharging}
\bibinfo{author}{Truzzolillo, D.} \emph{et~al.}
\newblock \bibinfo{title}{Overcharging and reentrant condensation of
  thermoresponsive ionic microgels}.
\newblock \emph{\bibinfo{journal}{Soft Matter}} \textbf{\bibinfo{volume}{14}},
  \bibinfo{pages}{4110--4125} (\bibinfo{year}{2018}).

\bibitem{acciaro:2011}
\bibinfo{author}{Acciaro, R.}, \bibinfo{author}{Gilányi, T.} \&
  \bibinfo{author}{Varga, I.}
\newblock \bibinfo{title}{Preparation of monodisperse
  poly(n-isopropylacrylamide) microgel particles with homogenous cross-link
  density distribution}.
\newblock \emph{\bibinfo{journal}{Langmuir}} \textbf{\bibinfo{volume}{27}},
  \bibinfo{pages}{7917--7925} (\bibinfo{year}{2011}).

\bibitem{tiwari:2014}
\bibinfo{author}{Tiwari, R.} \emph{et~al.}
\newblock \bibinfo{title}{A versatile synthesis platform to prepare uniform,
  highly functional microgels via click-type functionalization of latex
  particles}.
\newblock \emph{\bibinfo{journal}{Macromolecules}}
  \textbf{\bibinfo{volume}{47}}, \bibinfo{pages}{2257--2267}
  (\bibinfo{year}{2014}).

\bibitem{wei:2016}
\bibinfo{author}{Wei, J.}, \bibinfo{author}{Li, Y.} \& \bibinfo{author}{Ngai,
  T.}
\newblock \bibinfo{title}{Tailor-made microgel particles: Synthesis and
  characterization}.
\newblock \emph{\bibinfo{journal}{Colloid. Surface A}}
  \textbf{\bibinfo{volume}{489}}, \bibinfo{pages}{122 -- 127}
  (\bibinfo{year}{2016}).

\bibitem{mueller:2018}
\bibinfo{author}{Mueller, E.} \emph{et~al.}
\newblock \bibinfo{title}{Dynamically cross-linked self-assembled
  thermoresponsive microgels with homogeneous internal structures}.
\newblock \emph{\bibinfo{journal}{Langmuir}} \textbf{\bibinfo{volume}{34}},
  \bibinfo{pages}{1601--1612} (\bibinfo{year}{2018}).

\bibitem{batchelor:77}
\bibinfo{author}{Batchelor, G.}
\newblock \bibinfo{title}{The effect of brownian motion on the bulk stress in a
  suspension of spherical particles}.
\newblock \emph{\bibinfo{journal}{J. Fluid Mech.}}
  \textbf{\bibinfo{volume}{83}}, \bibinfo{pages}{97--117}
  (\bibinfo{year}{1977}).

\bibitem{Crocker:96}
\bibinfo{author}{Crocker, J.~C.} \& \bibinfo{author}{Grier, D.~G.}
\newblock \bibinfo{title}{Methods of digital video microscopy for colloidal
  studies}.
\newblock \emph{\bibinfo{journal}{J. Colloid Interf. Sci.}}
  \textbf{\bibinfo{volume}{179}}, \bibinfo{pages}{298--310}
  (\bibinfo{year}{1996}).

\bibitem{Parola:15}
\bibinfo{author}{Parola, A.} \& \bibinfo{author}{Reatto, L.}
\newblock \bibinfo{title}{Depletion interaction between spheres of unequal size
  and demixing in binary mixtures of colloids}.
\newblock \emph{\bibinfo{journal}{Mol. Phys.}} \textbf{\bibinfo{volume}{113}},
  \bibinfo{pages}{2571--2582} (\bibinfo{year}{2015}).

\bibitem{jeromemanuscript}
\bibinfo{author}{Paloli, D.}, \bibinfo{author}{Crassous, J.~J.},
  \bibinfo{author}{Mohanty, P.~S.}, \bibinfo{author}{Zaccarelli, E.} \&
  \bibinfo{author}{Schurtenberger, P.}
\newblock \bibinfo{title}{Viscoelastic properties of dense thermoresponsive
  microgel disperions at the glass transition and far beyond}.
\newblock \bibinfo{note}{Manuscript in preparation}.

\bibitem{landau:58}
\bibinfo{author}{Landau, L.~D.} \& \bibinfo{author}{Lifshitz, E.~M.}
\newblock \emph{\bibinfo{title}{Quantum Mechanics: Non-relativistic Theory. V.
  3 of Course of Theoretical Physics}} (\bibinfo{publisher}{Pergamon Press},
  \bibinfo{year}{1958}).

\bibitem{Riest:12}
\bibinfo{author}{Riest, J.}, \bibinfo{author}{Mohanty, P.},
  \bibinfo{author}{Schurtenberger, P.} \& \bibinfo{author}{Likos, C.~N.}
\newblock \bibinfo{title}{Coarse-graining of ionic microgels: Theory and
  experiment}.
\newblock \emph{\bibinfo{journal}{Z. Phys. Chem.}}
  \textbf{\bibinfo{volume}{226}}, \bibinfo{pages}{711--735}
  (\bibinfo{year}{2012}).

\bibitem{grest1986molecular}
\bibinfo{author}{Grest, G.~S.} \& \bibinfo{author}{Kremer, K.}
\newblock \bibinfo{title}{Molecular dynamics simulation for polymers in the
  presence of a heat bath}.
\newblock \emph{\bibinfo{journal}{Phys. Rev. A}} \textbf{\bibinfo{volume}{33}},
  \bibinfo{pages}{3628} (\bibinfo{year}{1986}).

\bibitem{roux1995calculation}
\bibinfo{author}{Roux, B.}
\newblock \bibinfo{title}{The calculation of the potential of mean force using
  computer simulations}.
\newblock \emph{\bibinfo{journal}{Comput. Phys. Commun.}}
  \textbf{\bibinfo{volume}{91}}, \bibinfo{pages}{275--282}
  (\bibinfo{year}{1995}).

\bibitem{mladek2011pair}
\bibinfo{author}{Mladek, B.~M.} \& \bibinfo{author}{Frenkel, D.}
\newblock \bibinfo{title}{Pair interactions between complex mesoscopic
  particles from widom's particle-insertion method}.
\newblock \emph{\bibinfo{journal}{Soft Matter}} \textbf{\bibinfo{volume}{7}},
  \bibinfo{pages}{1450--1455} (\bibinfo{year}{2011}).

\bibitem{Allen}
\bibinfo{author}{Allen, M.~P.} \& \bibinfo{author}{Tildesley, D.~J.}
\newblock \emph{\bibinfo{title}{Computer simulation of liquids}}
  (\bibinfo{publisher}{Oxford university press}, \bibinfo{year}{2017}).

\bibitem{Josephson:16}
\bibinfo{author}{Josephson, L.~L.}, \bibinfo{author}{Furst, E.~M.} \&
  \bibinfo{author}{Galush, W.~J.}
\newblock \bibinfo{title}{Particle tracking microrheology of protein
  solutions}.
\newblock \emph{\bibinfo{journal}{J. Rheol.}} \textbf{\bibinfo{volume}{60}},
  \bibinfo{pages}{531--540} (\bibinfo{year}{2016}).

\end{thebibliography}

%----------------------------------------------------------------------------------------
%	END NOTES
%----------------------------------------------------------------------------------------
%

\section*{Acknowledgments} We thank Sofi N\"{o}jd for particle synthesis and Andrea Ninarello for discussions. MB and PS acknowledge financial support from the European Research Council (ERC-339678-COMPASS) and the Swedish Research Council (VR 2015-05426). NG, LR and EZ acknowledge support from the European Research Council (ERC Consolidator Grant 681597, MIMIC). 

\section*{Author contributions}
EZ and PS designed and supervised research. MB performed all experiments with help from MOR and JMM. MB, NG, LR and EZ performed simulations and modeling. 
All authors contributed to the interpretation and analysis of the data. MB, LR, EZ and PS wrote the manuscript with inputs from all other authors.

\section*{Additional Information}
Competing  interests:
The authors declare no competing  interests.

%----------------------------------------------------------------------------------------
%	FIGURES AND LEGENDS
%----------------------------------------------------------------------------------------
%

 \vspace{-5.0cm}
  
 \begin{table}[t!]
\centering
\caption{{\bf Summary of effective volume fractions for all samples at all temperatures.}}
\begin{tabular}{r|c|ccc}
							&one-component system		&\multicolumn{3}{c}{binary mixtures} \\
$T$ ($\degree$C)				&	$\phi_{\text{eff,c}} $		& $\phi_{\text{eff,d}}$	& $\phi_{\text{eff,d}}$	&$\phi_{\text{eff,d}}$ \\
\hline
15					& 0.26			& 0.18	& 0.26		& 0.30 \\
15					& 0.37			& 0.18	& 0.26		& 0.30 \\
15					& 0.49			& 0.18	& 0.26		& 0.30 \\
\hline
20					& 0.22			& 0.17	& 0.24		& 0.28 \\
20					& 0.315			& 0.17	& 0.24		& 0.28 \\
20					& 0.42			& 0.17	& 0.24		& 0.28 \\
\hline
25					& 0.19			& 0.16	& 0.23		& 0.26 \\
25					& 0.275			& 0.16	& 0.23		& 0.26 \\
25					& 0.37			& 0.16	& 0.23		& 0.26 \\
\hline
30					& 0.155			& 0.145	& 0.215		& 0.24 \\
30					& 0.22			& 0.145	& 0.215		& 0.24 \\
30					& 0.29			& 0.145	& 0.215		& 0.24 \\
\end{tabular}
\label{tab:phieff}
\end{table}

 \begin{table}[h]
\centering
\caption{{\bf Summary of all parameters for the MH model and depletion term as a function of temperature.} Hertzian strengths $U_{\text{core}}$, $U_{\text{mid}}$, $U_{\text{corona}}$ and $U_{\text{cc}}$ in $k_\text{B}T$ and lengths $\sigma_{\text{core}}, \sigma_{\text{mid}}, \sigma_{\text{corona}}$ and $\sigma_{\text{eff}}$ as used for colloid-colloid interactions. Hertzian strength $U_{\text{cd}}$ in $k_\text{B}T$ used to describe the colloid-depletant (non-additive) interaction in binary mixtures. }
\begin{tabular}{r|cccc|r|cccc}
$T$ ($\degree$C)				&	$U_{\text{core}}$		& $U_{\text{mid}}$	& $U_{\text{corona}}$ 		& $U_{\text{cc}}$	&$U_{\text{cd}}$&$\sigma_{\text{core}}$ &$\sigma_{\text{mid}}$  & $\sigma_{\text{corona}}$ & $\sigma_{\text{eff}}$ \\
\hline
15					& 10 000			& 4 000			& 3 300	& 400		& 80 & 0.7			& 0.85		& 0.925		& 1\\\
20					& 10 000			& 5 200			& 4 300	& 520		& 108 & 0.74		& 0.87		& 0.935		& 1 \\
25					& 10 000			& 6 400			& 5 300	 & 640		& 144 & 0.77		& 0.885		& 0.945		& 1\\
30					& 10 000			& 7 600			& 6 300	 & 760		& 180 & 0.83		& 0.915		& 0.960		& 1\\
\end{tabular}
\label{tab:MH}
\end{table} 
\clearpage

%----------------------------------------------------------------------------------------
%	SUPPLEMENTARY INFORMATION
%----------------------------------------------------------------------------------------
%

\newpage
\renewcommand{\thefigure}{S\arabic{figure}}
\setcounter{figure}{0}
\begin{centering}
{\large Supplementary Information for: A new look at effective interactions of microgel particles via temperature dependence and depletion effects\\}

\vspace{50px}

Bergman et al. \\
\end{centering}
\clearpage
\section*{{\large Supplementary Figures}}

\begin{figure}[h!]
\centering
 \includegraphics{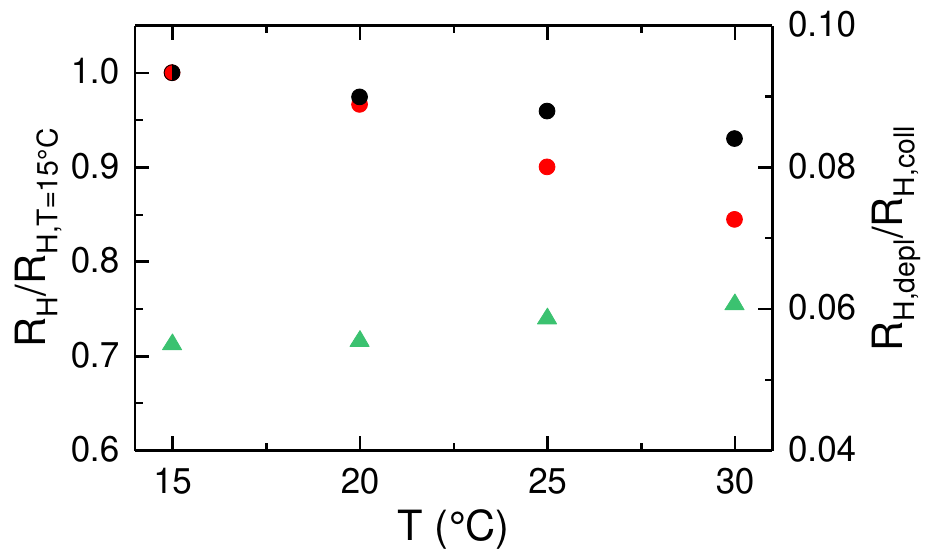}
  \caption{{\bf Swelling curves and size ratio.} Left axis: Relative swelling behaviour for colloid particles (red circles) for which $R_{\text{H},T=15\degree\text{C}} = 425$nm and depletant particles (black circles) for which $R_{\text{H},T=15\degree\text{C}} = 23$nm. Right axis: size ratio  $R_{\text{H}\text{,depletant}}/R_{H\text{,colloid}}$ (green triangles) as function of temperature.}
  \label{fig:swelling}
\end{figure}

 \begin{figure}[h!]
\centering
  \includegraphics{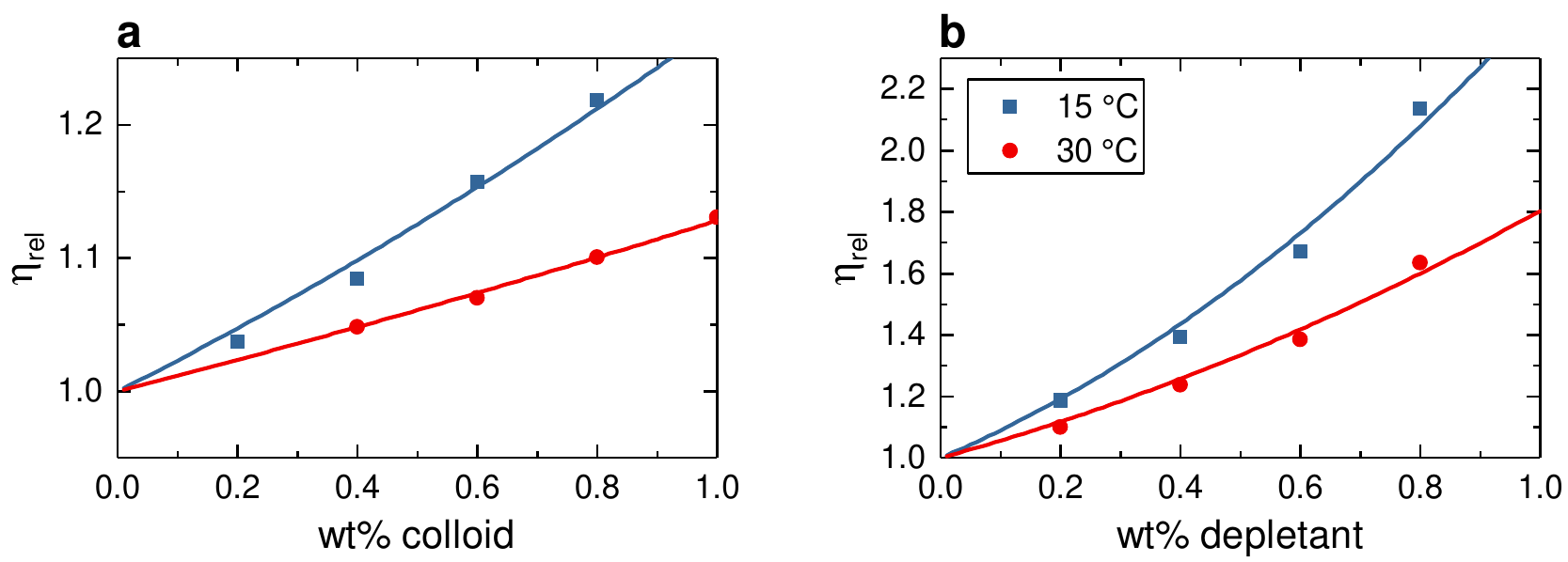}
 \caption{{\bf Experimental viscometry data acquired for {\bf a)} colloid only samples and {\bf b)} depletant only samples.} The relative viscosity was measured at 15 and 30$\degree$C (blue and red symbols) and fitted with the Batchelor equation (solid lines).}
  \label{fig:suppl-viscometry}
\end{figure}

 \begin{figure}[t!]
\centering
  \includegraphics[width=0.8\textwidth]{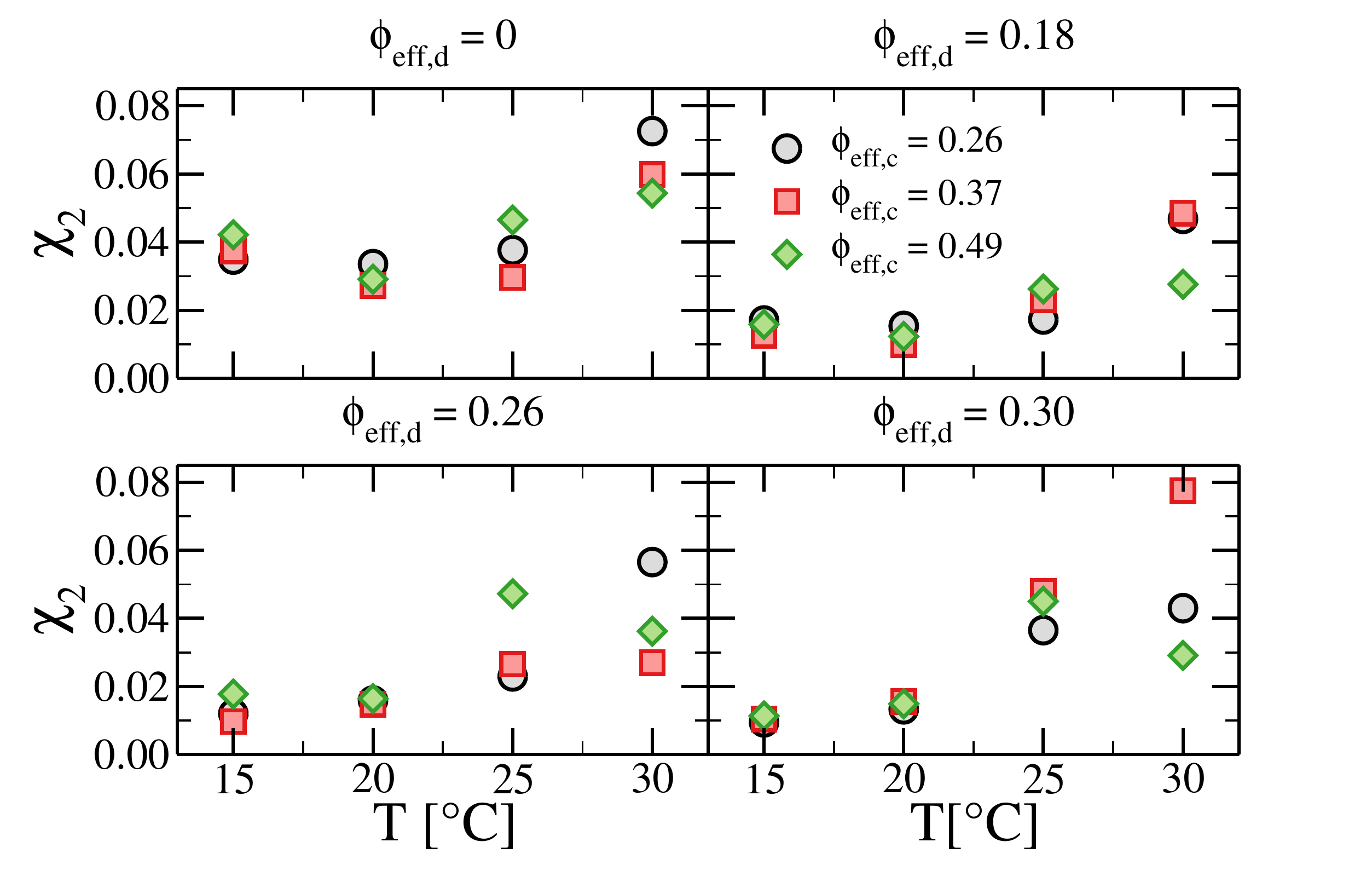}
 \caption{{\bf The average deviation $\chi^2$ between $g_{\rm sim}$ and $g_{\rm exp}$ for all the investigated state points.}}
  \label{fig:chi_2}
\end{figure}

 \begin{figure}[t!]
\centering
  \includegraphics[width=0.8\textwidth]{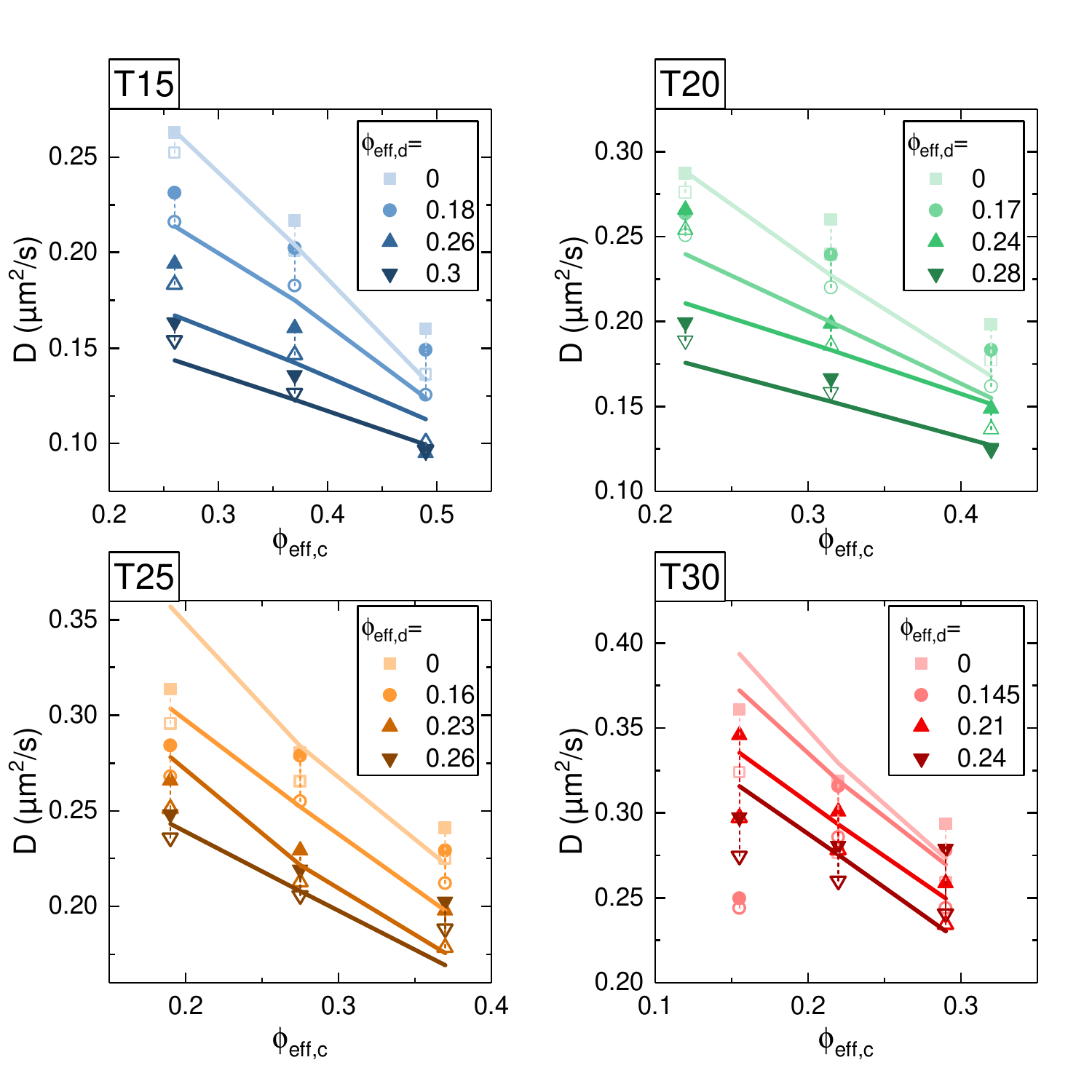}
 \caption{{\bf Self diffusion coefficients $D$ for each investigated state point plotted versus $\phi_{\text{eff,c}}$.} Symbols denote experimental data, lines represent simulated data. The diffusion coefficient taken from the slope of the MSD is shown as open symbols. The diffusion coefficient taken from the Van Hove self-correlations is shown as filled symbols.}
  \label{fig:diff_coeff}
\end{figure}

  \begin{figure}[t!]
\centering
 \includegraphics{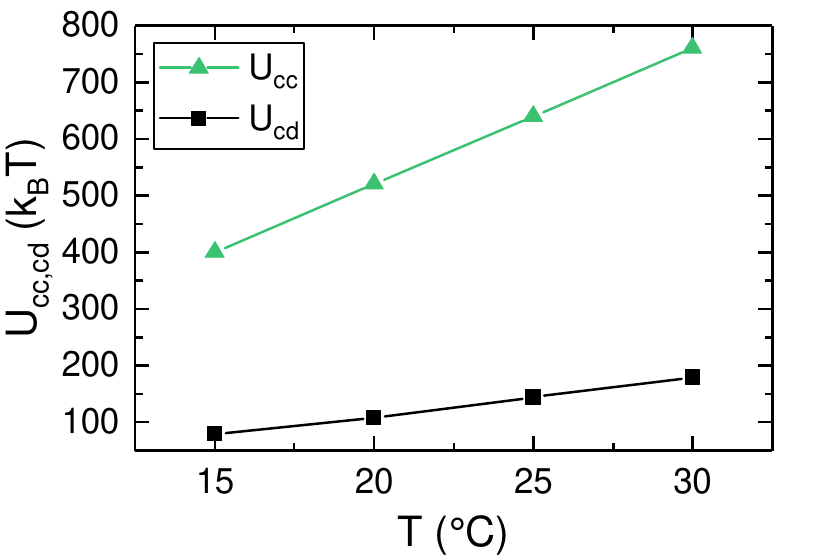}
 \caption{{\bf  Dependence of interactions strength on temperature.} Both $U_{\text{cc}}$ (green triangles) and $U_{\text{cd}}$ over temperature (black squares) are roughly linear.}
 \label{fig:Ucd-linear}
\end{figure}

%
% \begin{figure}[h]
%\centering
%  \includegraphics{Pick/t30forsuppl3inch}
%  \caption*{{\bf Supplementary Figure 4 Attempts to fit a typical experimental g(r) at 30$\degree$C.} Experimental data (symbols) shown for the colloid only sample with $\phi_{\text{eff,c}}=$0.29 at 30$\degree$C. Solid lines represent simulated $g(r)$s with increasing Hertzian repulsion.}
%	\label{fig:increasingH}
%\end{figure}

%\begin{figure}[h]
%\centering
  %\includegraphics{Pick-suppl/suppl-msds-v8-6inch}
 %\caption{{\bf Temperature dependence of {\bf a)} $g(r)$ and {\bf b)} MSD for two state points with the same effective packing fraction.} Experimental data for samples with $\phi_{\text{eff,c}}=$0.37 at two different temperatures ($T=$ 15 or 25 $\degree$C) corrected for size. Also shown in {\bf b)} is the predicted diffusion (dashed line).}
  %\label{fig:phi037}
%\end{figure}

 \begin{figure}[t!]
\centering
  \includegraphics{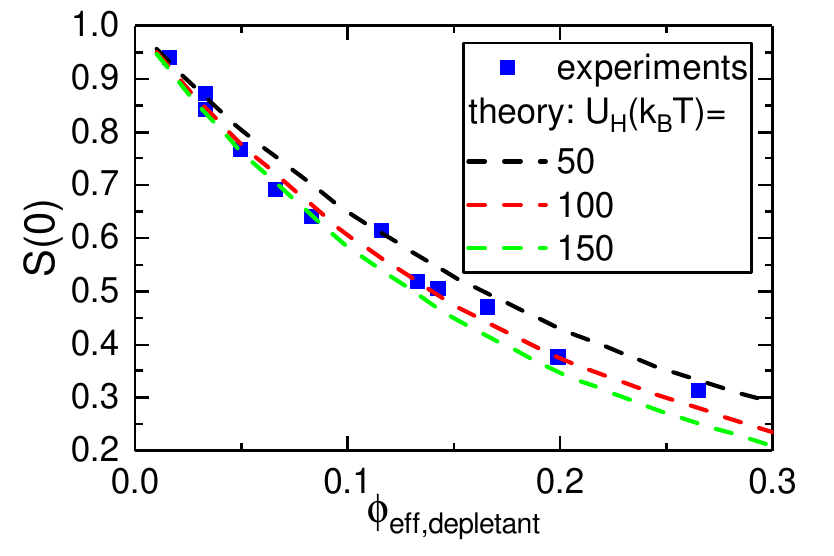}
  \caption{{\bf $S(0)$ data for depletants.} Experimental data (symbols) calculated from static light scattering measurements at 15$\degree$C are compared to theoretical values based on Hertzian potentials of varying interaction strength (dashed lines).}
  \label{fig:S0}
\end{figure}

  \begin{figure}[t!]
  \centering
  \includegraphics{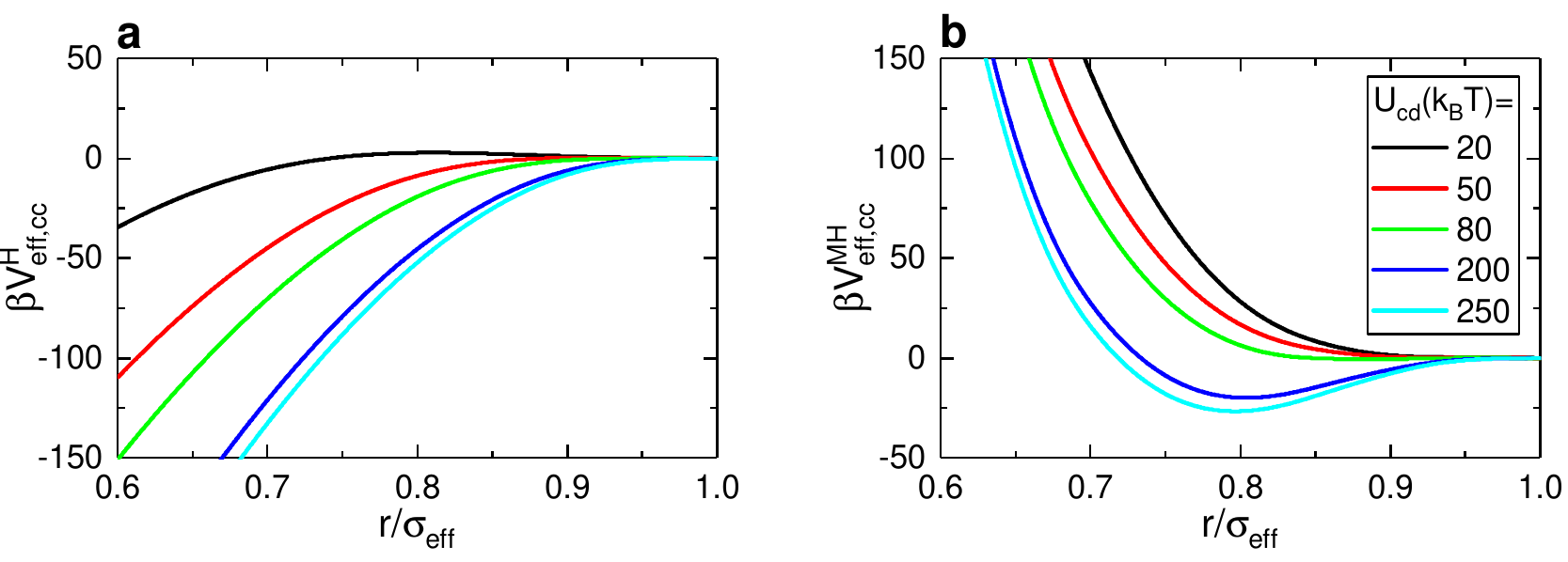}
  \caption{{\bf Variation of $U_{\text{cd}}$ in Hertzian and multi-Hertzian model.} 
  {\bf a)} Effect of $U_{\text{cd}}$ on $V^{\text{H}}_{\text{eff,cc}}$. 
  {\bf b)} Effect of varying $U_{\text{cd}}$ on $V^{\text{MH}}_{\text{eff,cc}}$. For the chosen strength (see text) $U_{\text{cd}}=80 k_\text{B}T$, the $V^{\text{MH}}_{\text{eff,cc}}$ displays a shallow negative minimum at $\sim -0.6k_\text{H}T$. Legend indicates interaction strength and is valid for both panels.}
  \label{fig:MH-pot}
  \end{figure}

   \begin{figure*}[t!]
\centering
  \includegraphics{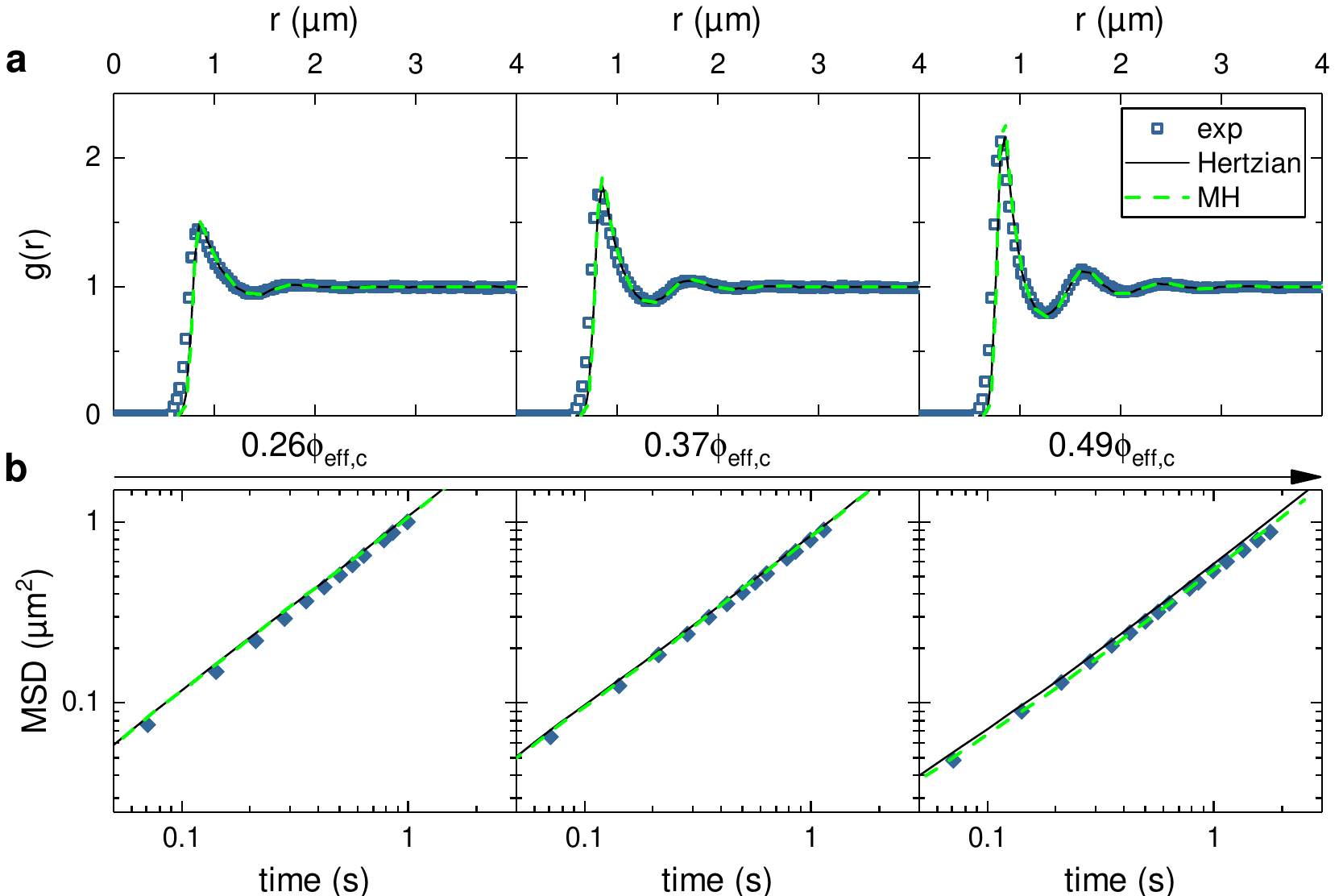}
  \caption{{\bf Experimental data at 15$\degree$C for one-component microgel system overlaid with simulated data from Hertzian and multi-Hertzian model.} 
  {\bf a)} Experimental $g(r)$s are plotted in the colored symbols with $\phi_{\text{eff,c}}=$0.26, 0.37 and 0.49 (left to right). Black solid lines show the calculated $g(r)$ from simulations based on the Hertzian potential. The overlapping green dotted lines show the calculated $g(r)$ from simulations based on the multi-Hertzian potential. 
  {\bf b)} Identical but for the MSD.}
  \label{fig:MHoverH}
\end{figure*} 

   \begin{figure*}[t!]
\centering
  \includegraphics[width=0.7\textwidth,clip]{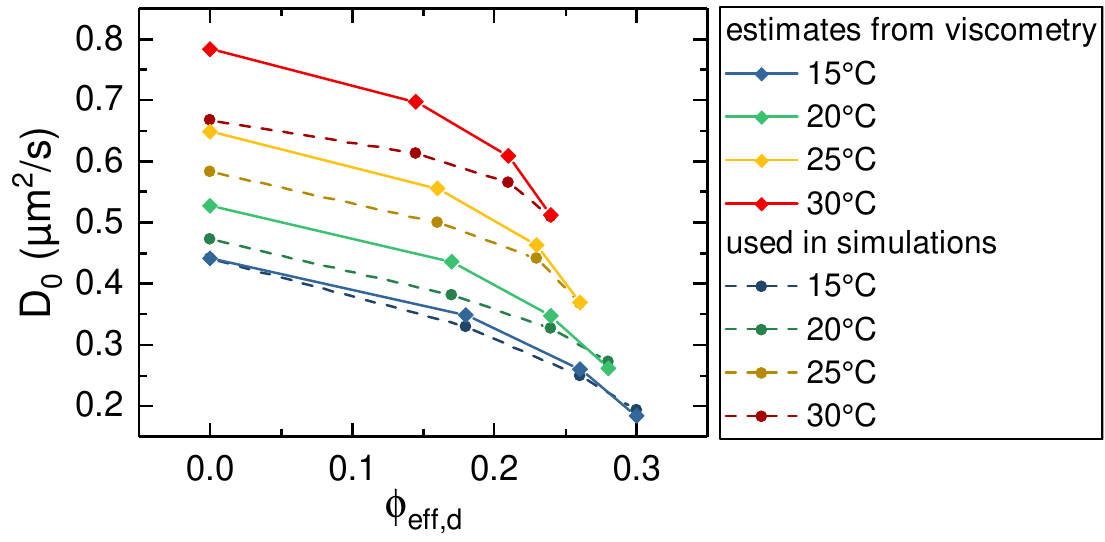}
  \caption{{\bf Zero-colloid limit self-diffusion coefficients $D_0$ for each investigated $T$ and $\phi_{\text{eff,d}}$.} Symbols connected by solid lines represent the estimates based on experimental data. Symbols connected by dashed lines represent the $D_0$ values used in simulations, which have been rescaled to $\phi_{\text{eff,d}}^{15^\circ C}$=0.3 for clarity.}
  \label{fig:D0}
\end{figure*}

\clearpage

\section*{Supplementary Notes}
%\newpage
{\bf Supplementary Note \ref{fig:swelling}: temperature dependent deswelling of the investigated microgels \\}

Supplementary Figure \ref{fig:swelling} shows the temperature dependent deswelling of the colloid (red symbols) and depletant microgels (black symbols). From this data we directly calculate the change in volume fraction at $T > 15\degree$C and the size ratio - which is shown on the right axis (green triangles). These parameters are directly used as inputs in the theoretical model.\\

%\newpage
{\bf Supplementary Note \ref{fig:suppl-viscometry}: bulk viscosity of colloid-only and depletant-only systems \\}

Supplementary Figure \ref{fig:suppl-viscometry} shows the relative viscosity for colloid-only samples (panel {\bf a}) and depletant only samples (panel {\bf b}) as function of weight percentage. Data was obtained at 15 and 30$\degree$C and fitted with the Batchelor equation. From this, we can obtain a shift factor $k$ (see also Materials and Methods) that allows us to move from wt\% to $\phi$ in the dilute limit. We use this conversion for our initial $\phi_{\text{eff}}$ guesses. In addition, relative viscosity data for the depletant-only samples is used to calculate the theoretical $D_0$ of colloid particles in the mixtures.\\ 

{\bf Supplementary Note \ref{fig:chi_2}: average deviations between experimental and numerical $g(r)$ \\}

Supplementary Figure~\ref{fig:chi_2} shows the average deviation between experimental and numerical $g(r)$ defined as $\chi^2 = \sum_i (g_{\rm sim}(r_i) - g_{\rm exp}(r_i))^2 / g_{\rm exp}^2(r_i)$. The state points where $\chi^2$ takes the smallest values are the ones at low $T$ and large $\phi_{\rm eff, d}$. Indeed, in addition to the systematic data noise associated to CLSM, at high $T$ or low depletant packing fractions, the lower spatial resolution in the $z$-direction and the rapid Brownian motion of the particles lead to a reduction in the peak height and a broadening of the $g(r)$ data at low values of $r$ to the left of the first peak, enhancing the average deviation.

{\bf Supplementary Note \ref{fig:diff_coeff}: comparison between the experimental and numerical self-diffusion coefficients \\}

We extracted the diffusion coefficient $D$ from the experimental data with two different approaches. First, we obtain the slope from the two-dimensional MSD versus lag time $\tau$ which corresponds to $4D$. This analysis heavily relies on the quality of the MSD and suffers from two main problems: for very short lag times  the particle displacement is difficult to resolve, while for very large lag times, the statistics are poor. Therefore, we complement this approach with that adopted in a recent work from Josephson et al. \cite{Josephson:16}. Hence, we calculate the one-dimensional Van Hove self-correlation for each specific lag time. The Van Hove self-correlation describes the probability of a particle moving $\Delta x$ in a given $\tau$. With enough statistics, such a probability should follow a Gaussian distribution and thus, the non-gaussian parameter $\alpha_2$ can be used to describe the shape of the curve.
In this way, we pinpoint the first lag time at which $\alpha_2$ indicates a normal distribution of the Van Hove function, i.e. the first lag time where we  get a reliable estimate of the diffusion coefficient with maximal statistics. Only the data at this particular $\tau$ is then used to calculate the diffusion coefficient using the Stokes-Einstein relation.  The self-diffusion coefficient from simulated data was determined using the slope of the MSD, as the statistics for the simulated MSD is very good up to large lag times.

The data in Supplementary Figure~\ref{fig:diff_coeff} show quite good agreement between experiments and simulations in the range 15-25 $^\circ$C. For $T=$30$^\circ$C the agreement worsens, probably because of the fast diffusion of the microgel particles.\\

{\bf Supplementary Note \ref{fig:Ucd-linear}: linear increase of the interaction strengths with temperature \\}

A posteriori, we find that the estimated values for $U_{\text{cc}}$ and $U_{\text{cd}}$ both increase almost linearly with temperature.
The  slope of $U_{\text{cd}}$ is slightly smaller than that found for $U_{\text{cc}}$, reflecting the fact that the small microgels are slightly softer (because of their lower crosslinker concentration) with respect to the large ones.\\

%{\bf Supplementary Note \ref{fig:phi037}: experimental data shows temperature dependence of microgel interactions}\\
%
%Supplementary Figure \ref{fig:phi037} shows the experimental structural and dynamic data for two distinctly different one-component microgel suspensions. The samples reach the same effective volume fraction at different temperatures, due to the temperature dependent deswelling of the microgels. Data have been rescaled to take into account of the variation of microgel diameters as well as of the change of bare diffusion coefficient $D_0$ at different temperatures. A higher structural correlation is observed for the sample at 25$\degree$C, which is associated to a smaller diffusion coefficient (relative to the bare diffusion). Thus it is evident that experimental data speak on their own,  suggesting that the interactions do change with temperature. The increase in correlation and reduced diffusion is indeed compatible with a stronger repulsion, as confirmed by the findings that the Hertzian model becomes \\ 

{\bf Supplementary Note \ref{fig:S0}: estimation of depletant-depletant interactions \\}

Supplementary Figure \ref{fig:S0} shows the measured data for the small wave vector limit $S(0)$ from static light scattering measurements at 15$\degree$C for the small depletant particles. $\phi_{\text{eff,d}}$ is calculated using the shift factor $k$ to go from wt\% to $\phi$. We compare the experimental data points with those calculated for a Hertzian potential using the Rogers-Young closure in the Ornstein-Zernike equation. From the comparison, we find that $U_{\text{dd}} \simeq 100k_\text{B}T$. The exact value is difficult to determine due to the spread in experimental data. However, the exact value is also not so important: whether the interaction strength lies between 50-150 $k_\text{B}T$ does not significantly affect our message. Indeed, the key point is that the interactions are very soft, and thus we can assume the depletant interactions to be ideal. In addition, the estimate of $U_{\text{dd}}$ can be used to estimate the cross-interactions between colloids and depletants. For additive interactions, any value of $U_{\text{dd}}$ would give a much higher value of $U_{\text{cd}}$ than what found in our model, suggesting a strong non-additivity of interactions in our soft-soft binary mixture.\\

%depicts the data used to estimate the soft repulsive Hertzian interaction between the small depletant particles. We obtained experimental values for the small wave vector limit $S(0)$ from static light scattering measurements at 15$\degree$C.   \\

{\bf Supplementary Note \ref{fig:MH-pot}: effect $U_{\text{cd}}$ on total interaction potential $V_{\text{eff,cc}}$ \\}

 Supplementary Figure \ref{fig:MH-pot} shows the effect of  the cross-interaction strength $U_{\text{cd}}$ on the total interaction potential $V_{\text{eff,cc}}$. We vary $U_{\text{cd}}$ from strongly non-additive at 20$k_\text{B}T$ to additive at 250$k_\text{B}T$. We find that even the smallest depletion interaction leads to a fully attractive potential if we use a Hertzian to describe colloid-colloid interactions (panel {\bf a}). If we add depletion attraction to the MH model, we recover a reasonable potential which reflects the stable experimental mixtures (panel {\bf b}).  \\

{\bf Supplementary Note \ref{fig:MHoverH}: comparison of Hertzian and MH model for one-component microgel systems\\}

Supplementary Figure \ref{fig:MHoverH} illustrates the comparison between Hertzian (solid lines) and MH (dashed lines) predictions for $g(r)$ (panel {\bf a}) and MSD (panel {\bf b}) of one-component microgel systems. Experimental data (symbols) are also shown as a reference.  At low $\phi_{\text{eff,c}}=0.26$ (left), overlap between the two models is perfect. However, small deviations are evident, particularly in the MSD, for the highest studied volume fraction $\phi_{\text{eff,c}}=0.49$ (right), where the MH model results are appreciably slower. Indeed, the inclusion of the core repulsion in the MH model is expected to provide a stronger contribution in situations where the microgels are packed tighter or forced together (e.g. due to compression or depletion effects). Thus, even for one-component microgel suspensions, it will be important not to neglect the internal architecture of the microgels to correctly describe the dynamics of the system at high densities.\\

{\bf Supplementary Note \ref{fig:D0}: comparison of the numerical and experimental zero-colloid limit self-diffusion coefficients\\}

We fix the numerical values of $D_0$ by comparing the long-time self-diffusion with the experimental data for each investigated $T$ and $\phi_{\text{eff,d}}$. The estimated $D_0$ values were kept fixed for all studied $\phi_{\text{eff,c}}$. The estimated $D_0$ are compared to the experimental viscometry values in Supplementary Figure~\ref{fig:D0}, finding a good agreement at large depletant concentrations and low temperatures, where hydrodynamic interactions are less important. \\

\end{document}